\documentclass[11pt,onecolumn]{article}%
\usepackage{amsmath}
\usepackage{amsfonts}
\usepackage{amssymb}
\usepackage{graphicx}
\usepackage{fullpage}%
\providecommand{\U}[1]{\protect\rule{.1in}{.1in}}
\newtheorem{theorem}{Theorem}

\newtheorem{definition}[theorem]{Definition}

\begin{document}

\title{Why Philosophers Should Care About Computational Complexity}
\author{Scott Aaronson\thanks{MIT. \ Email: aaronson@csail.mit.edu. \ \ This material
is based upon work supported by the National Science Foundation under Grant
No.\ 0844626. \ Also supported by a DARPA YFA grant, the Sloan Foundation, and
a TIBCO Chair.}}
\date{}
\maketitle

\begin{abstract}
One might think that, once we know something is computable, how
\textit{efficiently} it can be computed is a practical question with little
further philosophical importance. \ In this essay, I offer a detailed case
that one would be wrong. \ In particular, I argue that \textit{computational
complexity theory}---the field that studies the resources (such as time,
space, and randomness) needed to solve computational problems---leads to new
perspectives on the nature of mathematical knowledge, the strong AI debate,
computationalism, the problem of logical omniscience, Hume's problem of
induction, Goodman's grue riddle, the foundations of quantum mechanics,
economic rationality, closed timelike curves, and several other topics of
philosophical interest. \ I end by discussing aspects of complexity theory
itself that could benefit from philosophical analysis.

\end{abstract}
\tableofcontents

\section{Introduction\label{INTRO}}

\begin{quotation}
\noindent\textit{The view that machines cannot give rise to surprises is due,
I believe, to a fallacy to which philosophers and mathematicians are
particularly subject. This is the assumption that as soon as a fact is
presented to a mind all consequences of that fact spring into the mind
simultaneously with it. \ It is a very useful assumption under many
circumstances, but one too easily forgets that it is false. ---Alan M. Turing
\cite{turing:ai}}
\end{quotation}

The theory of computing, created by Alan Turing, Alonzo Church, Kurt
G\"{o}del, and others in the 1930s, didn't only change civilization; it also
had a lasting impact on philosophy. \ Indeed, clarifying philosophical issues
was the original \textit{point} of their work; the technological payoffs only
came later! \ Today, it would be hard to imagine a serious discussion about
(say) the philosophy of mind, the foundations of mathematics, or the prospects
of machine intelligence that was uninformed by this revolution in human
knowledge three-quarters of a century ago.

However, as computers became widely available starting in the 1960s, computer
scientists increasingly came to see computability theory as not asking quite
the right questions. \ For almost \textit{all} the problems we actually want
to solve turn out to be computable in Turing's sense; the real question is
which problems are \textit{efficiently}\ or \textit{feasibly}\ computable.
\ The latter question gave rise to a new field, called computational
complexity theory (not to be confused with the \textquotedblleft
other\textquotedblright\ complexity theory, which studies complex systems such
as cellular automata). \ Since the 1970s, computational complexity theory has
witnessed some spectacular discoveries, which include $\mathsf{NP}%
$-completeness, public-key cryptography, new types of mathematical proof (such
as probabilistic, interactive, and zero-knowledge proofs), and the theoretical
foundations of machine learning and quantum computation. \ To people who work
on these topics, the work of G\"{o}del\ and Turing may look in retrospect like
just a warmup to the \textquotedblleft big\textquotedblright\ questions about computation.

Because of this, I find it surprising that complexity theory has \textit{not}
influenced philosophy to anything like the extent computability theory has.
\ The question arises: \textit{why hasn't it?} \ Several possible answers
spring to mind: maybe computability theory just had richer philosophical
implications. \ (Though as we'll see, one can make a strong case for exactly
the opposite.) \ Maybe complexity has essentially the \textit{same}
philosophical implications as computability, and computability got there
first. \ Maybe outsiders are scared away from learning complexity theory by
the \textquotedblleft math barrier.\textquotedblright\ \ Maybe the explanation
is social: the world where G\"{o}del, Turing, Wittgenstein, and Russell
participated in the same intellectual conversation vanished with World War II;
after that, theoretical computer science came to be driven by technology\ and
lost touch with its philosophical origins. \ Maybe recent advances in
complexity theory simply haven't had enough time to enter philosophical consciousness.

However, I suspect that part of the answer is just \textit{complexity
theorists' failure to communicate} what they can add to philosophy's
conceptual arsenal. \ Hence this essay, whose modest goal is to help correct
that failure, by surveying some aspects of complexity theory that might
interest philosophers, as well as some philosophical problems that I think a
complexity perspective can clarify.

To forestall misunderstandings, let me add a note of humility before going
further. This essay will touch on many problems that philosophers have debated
for generations,\ such as strong AI, the problem of induction, the relation
between syntax and semantics, and the interpretation of quantum mechanics.
\ \textit{In none of these cases} will I claim that computational complexity
theory \textquotedblleft dissolves\textquotedblright\ the philosophical
problem---only that it contributes useful perspectives and insights. \ I'll
often explicitly mention philosophical puzzles that I think a complexity
analysis either leaves untouched or else introduces itself. \ But even where I
don't do so, one shouldn't presume that I think there are no such puzzles!
\ Indeed, one of my hopes for this essay is that computer scientists,
mathematicians, and other technical people who read it\ will come away with a
better appreciation for the subtlety of some of the problems considered in
modern analytic philosophy.\footnote{When I use the word \textquotedblleft
philosophy\textquotedblright\ in this essay, I'll \textit{mean} philosophy
within the analytic tradition. \ I don't understand Continental or Eastern
philosophy well enough to say whether they have any interesting connections
with computational complexity theory.}

\subsection{What This Essay \textit{Won't} Cover\label{WONT}}

I won't try to discuss every \textit{possible} connection between
computational complexity and philosophy, or even every connection that's
already been made. \ A small number of philosophers have long invoked
computational complexity ideas in their work; indeed, the \textquotedblleft
philpapers archive\textquotedblright\ lists $32$\ papers under the heading
Computational Complexity.\footnote{See
philpapers.org/browse/computational-complexity} \ The majority of those papers
prove theorems about the computational complexities of various logical
systems. \ Of the remaining papers, some use \textquotedblleft computational
complexity\textquotedblright\ in a different sense than I do---for example, to
encompass computability theory---and some invoke the \textit{concept} of
computational complexity, but no particular results from the \textit{field}
devoted to it. \ Perhaps the closest in spirit to this essay are the
interesting articles by Cherniak \cite{cherniak} and Morton \cite{morton}.
\ In addition, many writers have made some version of the observations in
Section \ref{AI}, about computational complexity and the\ Turing Test: see for
example Block \cite{block}, Parberry \cite{parberry}, Levesque \cite{levesque}%
, and Shieber \cite{shieber}.

In deciding which connections to include in this essay, I adopted the
following ground rules:

\begin{enumerate}
\item[(1)] The connection must involve a \textquotedblleft properly
philosophical\textquotedblright\ problem---for example, the justification for
induction or the nature of mathematical knowledge---and not just a technical
problem in logic or model theory.

\item[(2)] The connection must draw on \textit{specific insights} from the
field of computational complexity theory: not just the \textit{idea} of
complexity, or the \textit{fact} that there exist hard problems.
\end{enumerate}

There are many philosophically-interesting ideas in modern complexity theory
that this essay mentions only briefly or not at all. \ One example is
\textit{pseudorandom generators} (see Goldreich \cite{goldreich:prg}%
):\ functions that convert a short random \textquotedblleft
seed\textquotedblright\ into a long string of bits that, while not truly
random, is so \textquotedblleft random-looking\textquotedblright\ that no
efficient algorithm can detect any regularities in it. \ While pseudorandom
generators in this sense are not yet proved to exist,\footnote{The conjecture
that pseudorandom generators exist implies the $\mathsf{P}\neq\mathsf{NP}%
$\ conjecture (about which more later), but might be even stronger: the
converse implication is unknown.} there are many plausible candidates, and the
belief that at least some of the candidates work\ is central to modern
cryptography. \ (Section \ref{DRAWBACKS}\ will invoke the related concept of
pseudorandom \textit{functions}.) \ A second example is \textit{fully
homomorphic encryption}: an extremely exciting new class of methods, the first
of which was announced by Gentry \cite{gentry}\ in 2009, for performing
arbitrary computations on encrypted data \textit{without ever decrypting the
data}. \ The output of such a computation will look like meaningless gibberish
to the person who computed it, but it can nevertheless be understood (and even
recognized as the correct output) by someone who knows the decryption key.
\ What are the implications of pseudorandom generators for the foundations of
probability, or of fully homomorphic encryption for debates about the semantic
meaning of computations? \ I very much hope that this essay will inspire
others to tackle these and similar questions.

Outside of computational complexity, there are at least three other major
intersection points between philosophy and modern theoretical computer
science. \ The first one is the \textit{semantics of programming languages},
which has large and obvious connections to the philosophy of
language.\footnote{The \textit{Stanford Encyclopedia of Philosophy} entry on
\textquotedblleft The Philosophy of Computer Science,\textquotedblright%
\ plato.stanford.edu/entries/computer-science, devotes most of its space to
this connection.} \ The second is \textit{distributed systems theory}, which
provides both an application area and a rich source of examples for
philosophical work on reasoning about knowledge (see Fagin et al.\ \cite{fhmv}
and Stalnaker \cite{stalnaker}). \ The third is \textit{Kolmogorov complexity}
(see Li and Vit\'{a}nyi \cite{livitanyi}) which studies the \textit{length} of
the shortest computer program that achieves some functionality, disregarding
time, memory, and other resources used by the program.\footnote{A variant,
\textquotedblleft resource-bounded Kolmogorov complexity,\textquotedblright%
\ \textit{does} take time and memory into account, and is part of
computational complexity theory proper.}

In this essay, I won't discuss \textit{any} of these connections, except in
passing (for example, Section \ref{OMNI}\ touches on logics of knowledge\ in
the context of the \textquotedblleft logical omniscience
problem,\textquotedblright\ and Section \ref{PAC}\ touches on Kolmogorov
complexity in the context of PAC-learning). \ In defense of these omissions,
let me offer four excuses. \ First, these other connections fall outside my
stated topic. \ Second, they would make this essay even longer than it already
is. \ Third, I lack requisite background. \ And fourth, my impression is that
philosophers---at least \textit{some} philosophers---are already more aware of
these other connections than they are of the computational complexity
connections that I want to explain.

\section{Complexity 101\label{C101}}

Computational complexity theory is a huge, sprawling field; naturally this
essay will only touch\ on small parts of it. \ Readers who want to delve
deeper into the subject are urged to consult one of the many outstanding
textbooks, such as those of Sipser \cite{sipser:book}, Papadimitriou
\cite{papa:book}, Moore and Mertens \cite{mooremertens}, Goldreich
\cite{goldreich:book}, or Arora and Barak \cite{arorabarak}; or survey
articles by Wigderson \cite{wigderson:survey1,wigderson:survey2}, Fortnow and
Homer \cite{fortnowhomer}, or Stockmeyer \cite{stockmeyer:survey}.

One might think that, once we know something is \textit{computable}, whether
it takes $10$ seconds or $20$ seconds to compute is obviously the concern of
engineers rather than philosophers. \ But that conclusion would \textit{not}
be so obvious, if the question were one of $10$ seconds versus $10^{10^{10}}$
seconds! \ And indeed, in complexity theory, the quantitative gaps we care
about are usually so vast that one has to consider them qualitative gaps as
well. \ Think, for example, of the difference between reading a $400$-page
book and reading \textit{every possible} such book,\ or between writing down a
thousand-digit number and counting to that number.

More precisely, complexity theory asks the question: how do the resources
needed to solve a problem scale with some measure $n$ of the problem
size:\ \textquotedblleft reasonably\textquotedblright\ (like $n$\ or $n^{2}$,
say), or \textquotedblleft unreasonably\textquotedblright\ (like $2^{n}$\ or
$n!$)? \ As an example, two $n$-digit integers can be multiplied using $\sim
n^{2}$ computational steps (by the grade-school method),\ or even $\sim n\log
n\log\log n$\ steps (by more advanced methods \cite{schonhagestrassen}).
\ Either method is considered efficient. \ By contrast, the fastest known
method for the reverse operation---\textit{factoring} an $n$-digit integer
into primes---uses $\sim2^{n^{1/3}}$\ steps, which is considered
inefficient.\footnote{This method is called the \textit{number field sieve},
and the quoted running time depends on plausible but unproved conjectures in
number theory. \ The best \textit{proven} running time is $\sim2^{\sqrt{n}}$.
\ Both of these represent nontrivial improvements over the na\"{\i}ve method
of trying all possible divisors, which takes $\sim2^{n}$ steps. \ See
Pomerance \cite{pomerance}\ for a good survey of factoring algorithms.}
\ Famously, this conjectured gap between the inherent difficulties of
multiplying and factoring is the basis for most of the cryptography currently
used on the Internet.

Theoretical computer scientists generally call an algorithm \textquotedblleft
efficient\textquotedblright\ if its running time can be upper-bounded by any
polynomial function of $n$, and \textquotedblleft
inefficient\textquotedblright\ if its running time can be lower-bounded by any
exponential function of $n$.\footnote{In some contexts, \textquotedblleft
exponential\textquotedblright\ means $c^{n}$\ for some constant $c>1$, but in
most complexity-theoretic contexts it can also mean $c^{n^{d}}$\ for constants
$c>1$\ and $d>0$.} \ These criteria have the great advantage of theoretical
convenience. \ While the exact complexity of a problem might depend on
\textquotedblleft low-level encoding details,\textquotedblright\ such as
whether our Turing machine has one or two memory tapes, or how the inputs are
encoded as binary strings, where a problem falls on the polynomial/exponential
dichotomy can be shown to be independent of almost all such
choices.\footnote{This is not to say that \textit{no} details of the
computational model matter: for example, some problems are known to be
solvable in polynomial time on a quantum computer, but \textit{not} known to
be solvable in polynomial time on a classical computer! \ But in my view, the
fact that the polynomial/exponential distinction can \textquotedblleft
notice\textquotedblright\ a modelling choice of this magnitude is a feature of
the distinction, not a bug.} \ Equally important are the \textit{closure
properties} of polynomial and exponential time: a polynomial-time algorithm
that calls a polynomial-time subroutine still yields an overall
polynomial-time algorithm, while a polynomial-time algorithm that calls an
exponential-time subroutine (or vice versa) yields an exponential-time
algorithm. \ There are also more sophisticated reasons why theoretical
computer scientists focus on polynomial time (rather than, say, $n^{2}$\ time
or $n^{\log n}$\ time); we'll explore some of those reasons in Section
\ref{COBHAM}.

The polynomial/exponential distinction is open to obvious objections: an
algorithm that took $1.00000001^{n}$\ steps would be much faster in practice
than an algorithm that took $n^{10000}$\ steps! \ Furthermore, there are many
growth rates that fall between polynomial and exponential, such as $n^{\log
n}$\ and $2^{2^{\sqrt{\log n}}}$. \ But empirically, polynomial-time
\textit{turned out} to correspond to \textquotedblleft efficient in
practice,\textquotedblright\ and exponential-time to \textquotedblleft
inefficient in practice,\textquotedblright\ so often that complexity theorists
became comfortable making the identification. \ \textit{Why} the
identification works is an interesting question in its own right, one to which
we will return in Section \ref{CONC}.

\textit{A priori}, insisting that programs terminate after reasonable amounts
of time, that they use reasonable amounts of memory, etc.\ might sound like
relatively-minor amendments to Turing's notion of computation. \ In practice,
though, these requirements lead to a theory with a completely different
character than computability theory. \ Firstly, complexity has much closer
connections with the \textit{sciences}: it lets us pose questions about (for
example)\ evolution, quantum mechanics, statistical physics, economics, or
human language acquisition that would be meaningless from a computability
standpoint (since \textit{all} the relevant problems are computable).
\ Complexity also differs from computability in the diversity of mathematical
\textit{techniques} used: while initially complexity (like computability) drew
mostly on mathematical logic, today it draws on probability, number theory,
combinatorics, representation theory, Fourier analysis, and nearly every other
subject about which yellow books are written. \ Of course, this contributes
not only to complexity theory's depth but also to its perceived inaccessibility.

In this essay, I'll argue that complexity theory has direct relevance to major
issues in philosophy, including syntax and semantics, the problem of
induction, and the interpretation of quantum mechanics. \ Or that, at least,
whether complexity theory \textit{does or does not} have such relevance is an
important question for philosophy! \ My personal view is that complexity will
ultimately prove \textit{more} relevant to philosophy than computability was,
precisely because of the rich connections with the sciences mentioned earlier.

\section{The Relevance of Polynomial Time\label{CTP}}

Anyone who doubts the importance of the polynomial/exponential
distinction\ needs only ponder how many basic intuitions in math, science, and
philosophy already implicitly rely on that distinction. \ In this section I'll
give three examples.

\subsection{The Entscheidungsproblem Revisited\label{ENTSCH}}

The \textit{Entscheidungsproblem}\ was the dream, enunciated by David Hilbert
in the 1920s, of designing a mechanical procedure to determine the truth or
falsehood of any well-formed mathematical statement. \ According to the usual
story, Hilbert's dream was irrevocably destroyed by the work of G\"{o}del,
Church,\ and Turing in the 1930s. \ First, the Incompleteness Theorem showed
that no recursively-axiomatizable formal system can encode \textit{all and
only} the true mathematical statements.\ \ Second, Church's and Turing's
results showed that, even if we settle for an incomplete system $F$, there is
\textit{still} no mechanical procedure to sort mathematical statements into
the three categories \textquotedblleft provable in $F$,\textquotedblright%
\ \textquotedblleft disprovable in $F$,\textquotedblright\ and
\textquotedblleft undecidable in $F$.\textquotedblright

However, there is a catch in the above story, which was first pointed out by
G\"{o}del\ himself, in a 1956 letter to John von Neumann that has become
famous in theoretical computer science since its rediscovery in the 1980s (see
Sipser \cite{sipser:pnp} for an English translation). \ Given a formal system
$F$ (such as Zermelo-Fraenkel set theory), G\"{o}del wrote, consider the
problem of deciding whether a mathematical statement $S$\ has a proof in $F$
\textit{with }$n$\textit{ symbols or fewer. \ }Unlike Hilbert's original
problem, this \textquotedblleft truncated
Entscheidungsproblem\textquotedblright\ is clearly decidable. \ For, if
nothing else, we could always just program a computer to search through all
$2^{n}$\ possible bit-strings with $n$ symbols, and check whether any of them
encodes a valid $F$-proof of $S$. \ The issue is \textquotedblleft
merely\textquotedblright\ that this approach takes an astronomical amount of
time: if $n=1000$ (say), then the universe will have degenerated into black
holes and radiation long before a computer can check $2^{1000}$\ proofs!

But as G\"{o}del\ also pointed out, it's far from obvious how to
\textit{prove} that there isn't a much better approach: an approach that would
avoid brute-force search, and find proofs of size $n$ in time polynomial in
$n$. \ Furthermore:

\begin{quotation}
\noindent If there actually were a machine with [running time] $\sim Kn$ (or
even only with $\sim Kn^{2}$) [for some constant $K$ independent of $n$], this
would have consequences of the greatest magnitude. \ That is to say, it would
clearly indicate that, despite the unsolvability of the Entscheidungsproblem,
the mental effort of the mathematician in the case of yes-or-no questions
could be completely [added in a footnote: apart from the postulation of
axioms] replaced by machines. \ One would indeed have to simply select an $n$
so large that, if the machine yields no result, there would then also be no
reason to think further about the problem.
\end{quotation}

If we replace the \textquotedblleft$\sim Kn$ or $\sim Kn^{2}$%
\textquotedblright\ in\ G\"{o}del's challenge by $\sim Kn^{c}$\ for an
\textit{arbitrary} constant $c$, then we get precisely what computer science
now knows as the $\mathsf{P}\ $versus $\mathsf{NP}$ problem. \ Here
$\mathsf{P}$\ (Polynomial-Time) is, roughly speaking, the class of all
computational problems that are solvable by a polynomial-time algorithm.
\ Meanwhile, $\mathsf{NP}$\ (Nondeterministic Polynomial-Time) is the class of
computational problems for which a solution can be \textit{recognized} in
polynomial time, even though a solution might be very hard to
find.\footnote{Contrary to a common misconception, $\mathsf{NP}$\ does
\textit{not} stand for \textquotedblleft Non-Polynomial\textquotedblright!
\ There \textit{are} computational problems that are \textit{known} to require
more than polynomial time (see Section \ref{TIME}), but the $\mathsf{NP}%
$\ problems are not among those. \ Indeed, the classes $\mathsf{NP}$\ and
\textquotedblleft Non-Polynomial\textquotedblright\ have a nonempty
intersection exactly if $\mathsf{P}\neq\mathsf{NP}$.
\par
For detailed definitions of $\mathsf{P}$, $\mathsf{NP}$, and several hundred
other complexity classes, see my Complexity Zoo website:
www.complexityzoo.com.} \ (Think, for example, of factoring a large number, or
solving a jigsaw or Sudoku puzzle.) \ Clearly $\mathsf{P}\subseteq\mathsf{NP}%
$, so the question is whether the inclusion is strict. \ If $\mathsf{P}%
=\mathsf{NP}$, then the ability to \textit{check} the solutions to puzzles
efficiently would imply the ability to \textit{find} solutions efficiently.
\ An analogy would be if anyone able to \textit{appreciate} a great symphony
could also compose one themselves!

Given the intuitive implausibility of such a scenario, essentially all
complexity theorists proceed (reasonably, in my opinion) on the assumption
that\ $\mathsf{P}\neq\mathsf{NP}$, even if they publicly claim open-mindedness
about the question. \ Proving or disproving $\mathsf{P}\neq\mathsf{NP}$\ is
one of the seven million-dollar Clay Millennium Prize Problems\footnote{For
more information see www.claymath.org/millennium/P\_vs\_NP/
\par
My own view is that $\mathsf{P}\ $versus $\mathsf{NP}$\ is manifestly the
\textit{most important} of the seven problems! \ For if $\mathsf{P}%
=\mathsf{NP}$, then by G\"{o}del's argument, there is an excellent chance that
we could program our computers to solve the other six problems as well.}
(alongside the Riemann Hypothesis, the Poincar\'{e} Conjecture proved in
$2002$ by Perelman, etc.), which should give some indication of the problem's
difficulty.\footnote{One might ask: can we \textit{explain} what makes the
$\mathsf{P}\neq\mathsf{NP}$\ problem so hard, rather than just pointing out
that many smart people have tried to solve it and failed? \ After four decades
of research, we \textit{do} have partial explanations for the problem's
difficulty, in the form of formal \textquotedblleft barriers\textquotedblright%
\ that rule out large classes of proof techniques. \ Three barriers identified
so far are \textit{relativization} \cite{bgs} (which rules out diagonalization
and other techniques with a \textquotedblleft computability\textquotedblright%
\ flavor), \textit{algebrization} \cite{awig} (which rules out diagonalization
even when combined with the main non-relativizing techniques known today), and
\textit{natural proofs} \cite{rr} (which shows that many \textquotedblleft
combinatorial\textquotedblright\ techniques, if they worked, could be turned
around to get faster algorithms to distinguish random from pseudorandom
functions).}

Now return to the problem of whether a mathematical statement $S$ has a proof
with $n$ symbols or fewer,\ in some formal system $F$. \ A suitable
formalization of this problem is easily seen to be in $\mathsf{NP}$. \ For
\textit{finding} a proof might be intractable, but if we're \textit{given} a
purported proof, we can certainly check in time polynomial in $n$ whether each
line of the proof follows by a simple logical manipulation of previous lines.
\ Indeed, this problem turns out to be $\mathsf{NP}$\textit{-complete}, which
means that it belongs to an enormous class of $\mathsf{NP}$\ problems, first
identified in the 1970s, that \textquotedblleft capture the entire difficulty
of $\mathsf{NP}$.\textquotedblright\ \ A few other examples of $\mathsf{NP}%
$-complete problems are Sudoku and jigsaw puzzles, the Traveling Salesperson
Problem, and the satisfiability problem for propositional
formulas.\footnote{\label{FACTOR}By contrast, and contrary to a common
misconception, there is strong evidence that factoring integers is
\textit{not} $\mathsf{NP}$-complete. \ It is known that if $\mathsf{P}%
\neq\mathsf{NP}$, then there are $\mathsf{NP}$\ problems that are neither in
$\mathsf{P}$\ nor $\mathsf{NP}$-complete \cite{ladner},\ and factoring is one
candidate for such a problem. \ This point will become relevant when we
discuss quantum computing.} \ Asking whether $\mathsf{P}=\mathsf{NP}$\ is
equivalent to asking whether \textit{any} $\mathsf{NP}$-complete\ problem can
be solved in polynomial time, and is also equivalent to asking whether
\textit{all} of them can be.

In modern terms, then, G\"{o}del is saying that if $\mathsf{P}=\mathsf{NP}$,
then whenever a theorem had a proof of reasonable length, we could
\textit{find} that proof in a reasonable amount of time. \ In such a
situation, we might say that \textquotedblleft for all practical
purposes,\textquotedblright\ Hilbert's dream of mechanizing mathematics had
prevailed, despite the undecidability results of G\"{o}del, Church, and
Turing. \ If you accept this, then it seems fair to say that until
$\mathsf{P}$ versus $\mathsf{NP}$\ is solved, the story of Hilbert's
Entscheidungsproblem---its rise, its fall, and the consequences for
philosophy---is not yet over.

\subsection{Evolvability\label{EVOL}}

Creationists often claim that Darwinian evolution is as vacuous an explanation
for complex adaptations as\ \textquotedblleft a tornado assembling a 747
airplane as it passes through a junkyard.\textquotedblright\ \ Why is this
claim false? \ There are several related ways of answering the question, but
to me, one of the most illuminating is the following. \ In principle, one
\textit{could} see a 747 assemble itself in a tornado-prone junkyard---but
before that happened, one would need to wait for an expected number of
tornadoes that grew \textit{exponentially} with the number of pieces of
self-assembling junk. \ (This is similar to how, in thermodynamics, $n$ gas
particles in a box \textit{will} eventually congregate themselves in one
corner of the box, but only after $\sim c^{n}$ time for some constant $c$.)
\ By contrast, evolutionary processes can often be observed in
simulations---and in some cases, even proved theoretically---to find
interesting solutions to optimization problems after a number of steps that
grows only \textit{polynomially} with the number of variables.

Interestingly, in a 1972 letter to Hao Wang (see \cite[p. 192]{wang}), Kurt
G\"{o}del expressed his own doubts about evolution as follows:

\begin{quotation}
\noindent I believe that mechanism in biology is a prejudice of our time which
will be disproved. \ In this case, one disproof, in my opinion, will consist
in a mathematical theorem to the effect that the formation within geological
time of a human body by the laws of physics (or any other laws of similar
nature), starting from a random distribution of the elementary particles and
the field, is as unlikely as the separation by chance of the atmosphere into
its components.
\end{quotation}

Personally, I see no reason to accept G\"{o}del's intuition on this subject
over the consensus of modern biology! \ But pay attention to G\"{o}del's
characteristically-careful phrasing. He does not ask whether evolution can
\textit{eventually} form a human body (for he knows that it can, given
exponential time); instead, he asks whether it can do so on a
\textquotedblleft merely\textquotedblright\ geological\textit{ }timescale.
\ Just as G\"{o}del's letter to von Neumann\ anticipated the $\mathsf{P}%
\ $versus $\mathsf{NP}$ problem, so G\"{o}del's letter to Wang might be said
to anticipate a recent effort, by the celebrated computer scientist Leslie
Valiant, to construct a quantitative \textquotedblleft theory of
evolvability\textquotedblright\ \cite{valiant:evol}. \ Building on Valiant's
earlier work in computational learning theory (discussed in Section
\ref{PAC}), evolvability tries to formalize and answer questions about the
\textit{speed} of evolution.\ \ For example: \textquotedblleft what sorts of
adaptive behaviors can evolve, with high probability, after only a polynomial
number of generations? \ what sorts of behaviors can be learned in polynomial
time, but \textit{not} via evolution?\textquotedblright\ \ While there are
some interesting early results, it should surprise no one that evolvability is
nowhere close to being able to calculate, from first principles, whether four
billion years is a \textquotedblleft reasonable\textquotedblright\ or
\textquotedblleft unreasonable\textquotedblright\ length of time for the human
brain to evolve out of the primordial soup.

As I see it, this difficulty reflects a general point about G\"{o}del's
\textquotedblleft evolvability\textquotedblright\ question. \ Namely, even
\textit{supposing} G\"{o}del was right, that the mechanistic worldview of
modern biology was \textquotedblleft as unlikely as the separation by chance
of the atmosphere into its components,\textquotedblright\ computational
complexity theory seems hopelessly far from being able to \textit{prove}
anything of the kind! \ In 1972, one could have argued that this merely
reflected the subject's newness: no one had thought terribly deeply yet about
how to prove \textit{lower bounds} on computation time. \ But by now, people
\textit{have} thought deeply about it, and have identified huge obstacles to
proving even\ such \textquotedblleft obvious\textquotedblright\ and
well-defined conjectures as $\mathsf{P}\neq\mathsf{NP}$.\footnote{Admittedly,
one might be able to prove that \textit{Darwinian natural selection} would
require exponential time to produce some functionality, without thereby
proving that \textit{any} algorithm would require exponential time.}
\ (Section \ref{AI} will make a related point, about the difficulty of proving
nontrivial lower bounds on the time or memory needed by a computer program to
pass the Turing Test.)

\subsection{Known Integers\label{INTEGERS}}

My last example of the philosophical relevance of the polynomial/exponential
distinction concerns the concept of \textquotedblleft
knowledge\textquotedblright\ in mathematics.\footnote{This section was
inspired by a question of A. Rupinski on the website \textit{MathOverflow}.
\ See
mathoverflow.net/questions/62925/philosophical-question-related-to-largest-known-primes/}
\ As of 2011, the \textquotedblleft largest known prime
number,\textquotedblright\ as reported by GIMPS (the Great Internet Mersenne
Prime Search),\footnote{www.mersenne.org} is $p:=2^{43112609}-1$. \ But on
reflection, what do we mean by saying that $p$ is \textquotedblleft
known\textquotedblright? \ Do we mean that, if we desired, we could literally
print out its decimal digits (using about $30,000$ pages)? \ That seems like
too restrictive a criterion. \ For, given a positive integer $k$ together with
a proof that $q=2^{k}-1$\ was prime, I doubt most mathematicians would
hesitate to call $q$ a \textquotedblleft known\textquotedblright\ prime,\ even
if $k$ were so large that printing out its decimal digits (or storing them in
a computer memory) were beyond the Earth's capacity. \ Should we call
$2^{2^{1000}}$\ an \textquotedblleft unknown power of $2$,\textquotedblright%
\ just because it has too many decimal digits to list before the Sun goes cold?

All that should \textit{really} matter, one feels, is that

\begin{enumerate}
\item[(a)] the expression `$2^{43112609}-1$'\ picks out a unique positive
integer, and

\item[(b)] that integer has been proven (in this case, via computer, of
course) to be prime.
\end{enumerate}

But wait! \ If those are the criteria, then why can't we immediately beat the
largest-known-prime record, like so?%
\[
p^{\prime}=\text{The first prime larger than }2^{43112609}-1.
\]
Clearly $p^{\prime}$ exists, it is unambiguously defined, and it is prime.
\ If we want, we can even write a program that is guaranteed to find
$p^{\prime}$ and output its decimal digits, using a number of steps that can
be upper-bounded \textit{a priori}.\footnote{For example, one could use
\textit{Chebyshev's Theorem} (also called \textit{Bertrand's Postulate}),
which says that for all $N>1$\ there\ exists a prime between $N$ and $2N$.}
\ Yet our intuition stubbornly insists that $2^{43112609}-1$ is a
\textquotedblleft known\textquotedblright\ prime in a sense that $p^{\prime}$
is not. \ Is there any principled basis for such a distinction?

The clearest basis that I can suggest is the following. \ We know an algorithm
that takes as input a positive integer $k$, and that outputs the decimal
digits of $p=2^{k}-1$\ \textit{using a number of steps that is
polynomial---indeed, linear---in the number of digits of }$p$. \ But we do not
know any similarly-efficient algorithm that provably outputs the first prime
larger than $2^{k}-1$.\footnote{\textit{Cram\'{e}r's Conjecture} states that
the spacing between two consecutive $n$-digit primes never exceeds $\sim
n^{2}$. \ This conjecture appears staggeringly difficult: even assuming the
Riemann Hypothesis, it is only known how to deduce the much weaker upper bound
$\sim n2^{n/2}$. \ But interestingly, if Cram\'{e}r's Conjecture is proved,
expressions like \textquotedblleft the first prime larger than $2^{k}%
-1$\textquotedblright\ will \textit{then} define \textquotedblleft known
primes\textquotedblright\ according to my criterion.}

\subsection{Summary}

The point of these examples was to illustrate that, beyond its utility for
theoretical computer science, the polynomial/exponential gap is also a fertile
territory for philosophy. \ I think of the polynomial/exponential gap as
occupying a \textquotedblleft middle ground\textquotedblright\ between two
other sorts of gaps: on the one hand, small quantitative gaps (such as the gap
between $n$\ steps and $2n$ steps); and on the other hand, the gap between a
finite number of steps and an infinite number. \ The trouble with small
quantitative gaps is that they are too sensitive to \textquotedblleft
mundane\textquotedblright\ modeling choices and the details of technology.
\ But the gap between finite and infinite has the opposite problem: it is
serenely \textit{in}sensitive to distinctions that we actually care about,
such as that between finding a solution and verifying it, or between classical
and quantum physics.\footnote{In particular, it is easy to check that the set
of \textit{computable} functions does not depend on whether we define
computability with respect to a classical or a quantum Turing machine, or a
deterministic or nondeterministic one. \ At most, these choices can change a
Turing machine's running time by an exponential factor, which is irrelevant
for computability theory.} \ The polynomial/exponential gap avoids both problems.

\section{Computational Complexity and the Turing Test\label{AI}}

\textit{Can a computer think?} \ For almost a century, discussions about this
question have often conflated two issues. \ The first is the \textquotedblleft
metaphysical\textquotedblright\ issue:

\begin{quotation}
\noindent Supposing a computer program passed the Turing Test (or as strong a
variant of the Turing Test as one wishes to define),\footnote{The Turing Test,
proposed by Alan Turing \cite{turing:ai}\ in 1950, is a test where a human
judge interacts with either another human or a computer conversation program,
by typing messages back and forth. \ The program \textquotedblleft
passes\textquotedblright\ the Test if the judge can't reliably distinguish the
program from the human interlocutor.
\par
By a \textquotedblleft strong variant\textquotedblright\ of the Turing Test, I
mean that besides the usual teletype conversation, one could add additional
tests requiring vision, hearing, touch, smell, speaking, handwriting, facial
expressions, dancing, playing sports and musical instruments, etc.---even
though many perfectly-intelligent \textit{humans} would then be unable to pass
the tests!} would we be right to ascribe to it \textquotedblleft
consciousness,\textquotedblright\ \textquotedblleft qualia,\textquotedblright%
\ \textquotedblleft aboutness,\textquotedblright\ \textquotedblleft
intentionality,\textquotedblright\ \textquotedblleft
subjectivity,\textquotedblright\ \textquotedblleft
personhood,\textquotedblright\ or whatever other charmed status we wish to
ascribe to other humans and to ourselves?
\end{quotation}

The second is the \textquotedblleft practical\textquotedblright\ issue:

\begin{quotation}
\noindent Could a computer program that passed (a strong version of) the
Turing Test actually be written? \ Is there some fundamental reason why it
couldn't be?
\end{quotation}

Of course, it was precisely in an attempt to separate these issues that Turing
proposed the Turing Test in the first place! \ But despite his efforts,\ a
familiar feature of anti-AI arguments to this day is that they first assert
AI's metaphysical impossibility, and then try to bolster that position with
claims about AI's practical difficulties. \ \textquotedblleft
Sure,\textquotedblright\ they say,\ \textquotedblleft a computer program might
mimic a few minutes of witty banter, but unlike a human being, it would never
show fear or anger or jealousy, or compose symphonies, or grow old, or fall in
love...\textquotedblright

The obvious followup question---and what if a program \textit{did} do all
those things?---is often left unasked, or else answered by listing more things
that a computer program could self-evidently never do. \ Because of this, I
suspect that many people who \textit{say} they consider AI a metaphysical
impossibility, really consider it only a practical impossibility: they simply
have not carried the requisite thought experiment far enough to see the
difference between the two.\footnote{One famous exception is John Searle
\cite{searle}, who has made it clear that, if (say) his best friend turned out
to be controlled by a microchip rather than a brain, then he would regard his
friend as never having been a person at all.} \ Incidentally, this is as
clear-cut a case as I know of where people would benefit from studying more philosophy!

Thus, the anti-AI arguments that interest me most have always been the ones
that target the practical issue from the outset,\ by proposing empirical
\textquotedblleft sword-in-the-stone tests\textquotedblright\ (in Daniel
Dennett's phrase \cite{dennett}) that it is claimed humans can pass but
computers cannot. \ The most famous such test\ is probably the one based on
G\"{o}del's Incompleteness Theorem, as proposed by John Lucas \cite{lucas} and
elaborated by Roger Penrose in his books \textit{The Emperor's New Mind}
\cite{penrose}\ and \textit{Shadows of the Mind} \cite{penrose:shadows}.

Briefly, Lucas and Penrose argued that, according to the Incompleteness
Theorem, one thing that a computer making deductions via fixed formal rules
can never do is to \textquotedblleft see\textquotedblright\ the consistency of
its own rules. \ Yet this, they assert, is something that human mathematicians
\textit{can} do, via some sort of intuitive perception of Platonic reality.
\ Therefore humans (or at least, human mathematicians!) can never be simulated
by machines.

Critics pointed out numerous holes in this argument,\footnote{See Dennett
\cite{dennett} and Chalmers \cite{chalmers}\ for example. \ To summarize:
\par
\begin{enumerate}
\item[(1)] Why should we assume a computer operates within a knowably-sound
formal system? \ If we grant a computer the same freedom to make occasional
mistakes that we grant humans, then the Incompleteness Theorem is no longer
relevant.
\par
\item[(2)] Why should we assume that human mathematicians have
\textquotedblleft direct perception of Platonic reality\textquotedblright?
\ Human mathematicians (such as Frege) have been wrong before about the
consistency of formal systems.
\par
\item[(3)] A computer could, of course, be programmed to output
\textquotedblleft I believe that formal system $F$ is
consistent\textquotedblright---and even to output answers to various followup
questions about \textit{why} it believes this.\ \ So in arguing that such
affirmations \textquotedblleft wouldn't really count\textquotedblright%
\ (because they wouldn't reflect \textquotedblleft true
understanding\textquotedblright),\ AI critics such as Lucas and Penrose are
forced to retreat from their vision of an empirical \textquotedblleft
sword-in-the-stone test,\textquotedblright\ and fall back on other,
unspecified criteria related to the AI's internal structure. \ But then
\textit{why put the sword in the stone in the first place?}
\end{enumerate}
} to which Penrose responded at length in \textit{Shadows of the Mind}, in my
opinion unconvincingly. \ However, even \textit{before} we analyze some
proposed sword-in-the-stone test, it seems to me that there is a much more
basic question. \ Namely, what does one even \textit{mean} in saying one has a
task that \textquotedblleft humans can perform but computers
cannot\textquotedblright?

\subsection{The Lookup-Table Argument\label{LOOKUP}}

There is a fundamental difficulty here, which was noticed by others in a
slightly different context \cite{block,parberry,levesque,shieber}. \ Let me
first explain the difficulty, and then discuss the difference between my
argument and the previous ones.

In practice, people judge each other to be conscious after interacting for a
very short time, perhaps as little as a few seconds. \ This suggests that we
can put a finite upper bound---to be generous, let us say $10^{20}$---on the
number of bits of information that two people $A$\ and $B$ would ever
realistically exchange, before $A$ had amassed enough evidence to conclude $B$
was conscious.\footnote{People interacting over the Internet, via email or
instant messages, regularly judge each other to be humans rather than
spam-bots after exchanging a much smaller number of bits! \ In any case,
cosmological considerations suggest an upper bound of roughly $10^{122}$\ bits
in any observable process \cite{bousso:vac}.} \ Now imagine a lookup table
that stores every possible history $H$\ of $A$ and $B$'s conversation, and
next to $H$, the action $f_{B}\left(  H\right)  $ that $B$\ \textit{would}
take next given that history. \ Of course, like Borges' Library of Babel, the
lookup table would consist almost entirely of meaningless nonsense, and it
would also be much too large to fit inside the observed universe. \ But all
that matters for us is that the lookup table would be \textit{finite}, by the
assumption that there is a finite upper bound on the conversation length.
\ This implies that the function $f_{B}$ is computable (indeed, it can be
recognized by a finite automaton!). \ From these simple considerations, we
conclude that if there \textit{is} a fundamental obstacle to computers passing
the Turing Test, then it is not to be found in computability
theory.\footnote{Some readers might notice a tension here: I explained in
Section \ref{C101} that complexity theorists care about the
\textit{asymptotic} behavior as the problem size $n$\ goes to infinity. \ So
why am I now saying that, for the purposes of the Turing Test, we should
restrict attention to finite values of $n$ such as $10^{20}$? \ There are two
answers to this question. \ The first is that, in contrast to mathematical
problems like the factoring problem or the halting problem, it is unclear
whether it even makes \textit{sense} to generalize the Turing Test to
arbitrary conversation lengths: for the Turing Test is defined in terms of
human beings, and human conversational capacity is finite. \ The second answer
is that, to whatever extent it \textit{does} make sense to generalize the
Turing Test to arbitrary conversation lengths $n$, I \textit{am} interested in
whether the asymptotic complexity of passing the test grows polynomially or
exponentially with $n$ (as the remainder of the section explains).}

In \textit{Shadows of the Mind} \cite[p. 83]{penrose:shadows}, Penrose
recognizes this problem, but gives a puzzling and unsatisfying response:

\begin{quotation}
\noindent One could equally well envisage computers that contain nothing but
lists of totally false mathematical `theorems,' or lists containing random
jumbles of truths and falsehoods. \ How are we to tell which computer to
trust? \ The arguments that I am trying to make here do not say that an
effective simulation of the output of conscious human activity (here
mathematics) is impossible, since purely by chance the computer might `happen'
to get it right---even without any understanding whatsoever. \ But the odds
against this are absurdly enormous, and the issues that are being addressed
here, namely how one decides \textit{which} mathematical statements are true
and which are false, are not even being touched...
\end{quotation}

The trouble with this response is that it amounts to a retreat from the
sword-in-the-stone test, back to murkier internal criteria. If, in the end, we
are going to have to look inside the computer anyway to determine whether it
truly \textquotedblleft understands\textquotedblright\ its answers,
\textit{then why not dispense with computability theory from the beginning?}
\ For computability theory only addresses whether or not Turing machines
\textit{exist} to solve various problems, and we have already seen that that
is not the relevant issue.

To my mind, there is \textit{one} direction that Penrose could take from this
point to avoid incoherence---though disappointingly, it is not the direction
he chooses. \ Namely, he could point out that, while the lookup table
\textquotedblleft works,\textquotedblright\ it requires computational
resources that grow exponentially\textit{ }with the length of the
conversation! \ This would lead to the following speculation:

\begin{quotation}
\noindent(*) \textit{Any} computer program that passed the Turing Test would
need to be exponentially-inefficient in the length of the test---as measured
in some resource such as time, memory usage, or the number of bits needed to
write the program down. \ In other words, the astronomical lookup table is
essentially the best one can do.\footnote{As Gil Kalai pointed out to me, one
could speculate instead that an efficient computer program \textit{exists} to
pass the Turing Test, but that \textit{finding} such a program would require
exponential computational resources. \ In that situation, the human brain
could indeed be simulated efficiently by a computer program, but maybe not by
a program that humans could ever \textit{write}!}
\end{quotation}

If true, speculation (*) would do what Penrose wants: it would imply that the
human brain can't even be \textit{simulated} by computer, within the resource
constraints of the observable universe. \ Furthermore, unlike the earlier
computability claim, (*) has the advantage of not being trivially false!

On the other hand, to put it mildly, (*) is not trivially \textit{true}
either. \ For AI proponents, the lack of compelling evidence for (*) is hardly
surprising.\ \ After all, if you believe that the brain \textit{itself} is
basically an efficient,\footnote{Here, by a Turing machine $M$ being
\textquotedblleft efficient,\textquotedblright\ we mean that $M$'s running
time, memory usage, and program size are modest enough that there is no real
problem of principle understanding how $M$ could be simulated by a classical
physical system consisting of $\thicksim10^{11}$ neurons and $\thicksim
10^{14}$\ synapses. \ For example, a Turing machine containing a lookup table
of size $10^{10^{20}}$\ would not be efficient in this sense.} classical
Turing machine, then you have a simple explanation for why no one has proved
that the brain can't be simulated by such a machine! \ However, complexity
theory also makes it clear that, \textit{even if we supposed (*) held}, there
would be little hope of \textit{proving} it in our current state of
mathematical knowledge. \ After all, we can't even prove plausible,
well-defined conjectures such as $\mathsf{P}\neq\mathsf{NP}$.

\subsection{Relation to Previous Work\label{PREVWORK}}

As mentioned before, I'm far from the first person to ask about the
\textit{computational resources} used in passing the Turing Test, and whether
they scale polynomially or exponentially with the conversation length. \ While
many writers ignore this crucial distinction, Block \cite{block}, Parberry
\cite{parberry}, Levesque \cite{levesque}, Shieber \cite{shieber}, and several
others all discussed it explicitly. \ The main difference is that the previous
discussions took place in the context of Searle's Chinese Room argument
\cite{searle}.

Briefly, Searle proposed a thought experiment---the details don't concern us
here---purporting to show that a computer program could pass the Turing Test,
even though the program manifestly lacked anything that a reasonable person
would call \textquotedblleft intelligence\textquotedblright\ or
\textquotedblleft understanding.\textquotedblright\ \ In response, many
critics said that Searle's argument was deeply misleading, because it
implicitly encouraged us to imagine a computer program that was
\textit{simplistic} in its internal operations---something like the giant
lookup table described in Section \ref{LOOKUP}. \ And while it was true, the
critics went on, that a giant lookup table wouldn't \textquotedblleft truly
understand\textquotedblright\ its responses, that point is also
\textit{irrelevant}. \ For the giant lookup table is a philosophical fiction
anyway: something that can't even fit in the observable universe! \ If we
instead imagine a \textit{compact, efficient} computer program passing the
Turing Test, then the situation changes drastically. \ For now, in order to
\textit{explain} how the program can be so compact and efficient, we'll need
to posit that the program includes representations of abstract concepts,
capacities for learning and reasoning, and all sorts of other internal
furniture that we would expect to find in a mind.

Personally, I find this response to Searle extremely interesting---since if
correct, it suggests that the distinction between polynomial and exponential
complexity has \textit{metaphysical} significance. \ According to this
response, an exponential-sized lookup table that passed the Turing Test would
not be sentient (or conscious, intelligent, self-aware, etc.), but a
polynomially-bounded program with exactly the same input/output behavior
\textit{would} be sentient. \ Furthermore, the latter program would be
sentient \textit{because} it was polynomially-bounded.

Yet, as much as that criterion for sentience flatters my complexity-theoretic
pride, I find myself reluctant to take a position on such a weighty matter.
\ My point, in Section \ref{LOOKUP}, was a simpler and (hopefully) less
controversial one:\ namely, that if you want to claim that passing the Turing
Test is \textit{flat-out impossible}, then like it or not, you \textit{must}
talk about complexity rather than just computability. \ In other words, the
previous writers \cite{block,parberry,levesque,shieber}\ and I are all
interested in the computational resources needed to pass a Turing Test of
length $n$, but for different reasons. \ Where others\ invoked complexity
considerations to argue with Searle about the metaphysical question, I'm
invoking them to argue with Penrose about the practical question.

\subsection{Can Humans Solve $\mathsf{NP}$-Complete Problems
Efficiently?\label{HUMANS}}

In that case, what can we actually \textit{say} about the practical question?
\ Are there any reasons to accept the claim I called (*)---the claim that
humans are \textit{not} efficiently simulable by Turing machines? \ In
considering this question, we're immediately led to some speculative
possibilities. \ So for example,\ \textit{if} it turned out that humans could
solve arbitrary instances of $\mathsf{NP}$-complete problems in polynomial
time, then that would certainly strong excellent empirical evidence for
(*).\footnote{And amusingly, if we could solve $\mathsf{NP}$-complete
problems, then we'd presumably find it much easier to prove that computers
\textit{couldn't} solve them!} \ However, despite occasional claims to the
contrary, I personally see no reason to believe that humans \textit{can} solve
$\mathsf{NP}$-complete problems in polynomial time, and excellent reasons to
believe the opposite.\footnote{Indeed, it is not even clear to me that we
should think of humans as being able to solve all $\mathsf{P}$\ problems
efficiently, let alone $\mathsf{NP}$-complete problems! \ Recall that
$\mathsf{P}$\ is the class of problems that \textit{are} solvable in
polynomial time by a deterministic Turing machine. \ Many problems are known
to belong to $\mathsf{P}$ for quite sophisticated reasons: two examples are
testing whether a number is prime (though not factoring it!) \cite{aks} and
testing whether a graph has a perfect matching. \ In principle, of course, a
human could laboriously run the polynomial-time algorithms for such problems
using pencil and paper. \ But is the use of pencil and paper legitimate, where
use of a computer would \textit{not} be? \ What is the computational power of
the \textquotedblleft unaided\textquotedblright\ human intellect? \ Recent
work of Drucker \cite{drucker}, which shows how to use a stock photography
collection to increase the \textquotedblleft effective
memory\textquotedblright\ available for mental calculations, provides a
fascinating empirical perspective on these questions.} \ Recall, for example,
that the integer factoring problem is in $\mathsf{NP}$. \ Thus, if humans
could solve $\mathsf{NP}$-complete problems, then presumably we ought to be
able to factor enormous numbers as well! \ But factoring does not exactly seem
like the most promising candidate for a sword-in-the-stone test: that is, a
task that's easy for humans but hard for computers. \ As far as anyone knows
today, factoring is hard for humans and (classical) computers\textit{ alike},
although with a definite advantage on the computers' side!

The basic point can hardly be stressed enough: when complexity theorists talk
about \textquotedblleft intractable\textquotedblright\ problems, they
generally mean mathematical problems that all our experience leads us to
believe are at least as hard for humans as for computers. \ This suggests
that, \textit{even if} humans were not efficiently simulable by Turing
machines, the \textquotedblleft direction\textquotedblright\ in which they
were hard to simulate would almost certainly be different from the directions
usually considered in complexity theory. \ I see two (hypothetical) ways this
could happen.

First, the tasks that humans were uniquely good at---like painting or writing
poetry---could be \textit{incomparable} with mathematical tasks like solving
$\mathsf{NP}$-complete problems, in the sense that neither was efficiently
reducible to the other. \ This would mean, in particular, that there could be
no polynomial-time algorithm even to \textit{recognize} great art or poetry
(since if such an algorithm existed, then the task of \textit{composing} great
art or poetry would be in $\mathsf{NP}$). \ Within complexity theory, it's
known that there exist pairs of problems that are incomparable in this sense.
\ As one plausible example, no one currently knows how to reduce the
simulation of quantum computers to the solution of $\mathsf{NP}$-complete
problems \textit{or} vice versa.

Second, humans could have the ability to solve interesting \textit{special
cases} of $\mathsf{NP}$-complete\ problems faster than any Turing machine.
\ So for example, even if computers were better than humans at factoring large
numbers or at solving randomly-generated Sudoku puzzles, humans might still be
better at search problems with \textquotedblleft higher-level
structure\textquotedblright\ or \textquotedblleft semantics,\textquotedblright%
\ such as proving Fermat's Last Theorem or (ironically) designing faster
computer algorithms. \ Indeed, even in limited domains such as puzzle-solving,
while computers can examine solutions millions of times faster, humans (for
now) are vastly better at noticing \textit{global patterns} or
\textit{symmetries} in the puzzle that make a solution either trivial or
impossible. \ As an amusing example, consider the \textit{Pigeonhole
Principle}, which says that $n+1$\ pigeons can't be placed into $n$ holes,
with at most one pigeon per hole. \ It's not hard to construct a propositional
Boolean formula $\varphi$\ that encodes the Pigeonhole Principle for some
fixed value of $n$ (say, $1000$). \ However, if you then feed $\varphi$\ to
current Boolean satisfiability algorithms, they'll assiduously set to work
trying out possibilities: \textquotedblleft let's see, if I put \textit{this}
pigeon here, and \textit{that} one there ... darn, it \textit{still} doesn't
work!\textquotedblright\ \ And they'll continue trying out possibilities for
an exponential number of steps, oblivious to the \textquotedblleft
global\textquotedblright\ reason why the goal can never be achieved. \ Indeed,
beginning in the 1980s, the field of \textit{proof complexity}---a close
cousin of computational complexity---has been able to show that large classes
of algorithms \textit{require} exponential time to prove the Pigeonhole
Principle and similar propositional tautologies (see Beame and Pitassi
\cite{beamepitassi}\ for a survey).

Still, if we want to build our sword-in-the-stone test on the ability to
detect \textquotedblleft higher-level patterns\textquotedblright\ in
combinatorial search problems, then the burden is on us to explain what we
\textit{mean} by higher-level patterns, and why we think that \textit{no}
polynomial-time Turing machine---even much more sophisticated ones than we can
imagine today---could ever detect those patterns as well. \ For an initial
attempt to understand $\mathsf{NP}$-complete\ problems\ from a cognitive
science perspective, see Baum \cite{baum}.

\subsection{Summary}

My conclusion is that, if you oppose the possibility of AI in principle, then either

\begin{enumerate}
\item[(i)] you can take the \textquotedblleft metaphysical
route\textquotedblright\ (as Searle \cite{searle}\ does with the Chinese
Room), conceding the possibility of a computer program passing every
conceivable empirical test for intelligence, but arguing that that isn't
enough, or

\item[(ii)] you can conjecture an \textit{astronomical lower bound on the
resources} needed either to run such a program or to write it in the first
place---but here there is little question of proof for the foreseeable future.
\end{enumerate}

\noindent Crucially, because of the lookup-table argument, one option you do
\textit{not} have is to assert the flat-out impossibility of a computer
program passing the Turing Test, with no mention of quantitative complexity bounds.

\section{The Problem of Logical Omniscience\label{OMNI}}

Giving a formal account of \textit{knowledge} is one of the central concerns
in modern analytic philosophy; the literature is too vast even to survey here
(though see Fagin et al.\ \cite{fhmv}\ for a computer-science-friendly
overview). \ Typically, formal accounts of knowledge\ involve conventional
\textquotedblleft logical\textquotedblright\ axioms, such as

\begin{itemize}
\item If you know $P$ and you know $Q$, then you also know $P\wedge Q$
\end{itemize}

\noindent supplemented by \textquotedblleft modal\textquotedblright\ axioms
having to do with knowledge itself, such as

\begin{itemize}
\item If you know $P$, then you also know that you know $P$

\item If you don't know $P$, then you know that you don't know $P$%
\footnote{Not surprisingly, this particular axiom has engendered controversy:
it leaves no possibility for Rumsfeldian \textquotedblleft unknown
unknowns.\textquotedblright}
\end{itemize}

While the details differ, what most formal accounts of knowledge have in
common is that they treat an agent's knowledge as \textit{closed} under the
application of various deduction rules like the ones above. \ In other words,
agents are considered \textit{logically omniscient}: if they know certain
facts, then they also know all possible logical consequences of those facts.

Sadly and obviously, no mortal being has ever attained or even approximated
this sort of omniscience (recall the Turing quote from the beginning of
Section \ref{INTRO}). \ So for example, I can know the rules of arithmetic
without knowing Fermat's Last Theorem, and I can know the rules of chess
without knowing whether White has a forced win. \ Furthermore, the difficulty
is \textit{not} (as sometimes claimed) limited to a few domains, such as
mathematics and games. \ As pointed out by Stalnaker \cite{stalnaker}, if we
assumed logical omniscience, then we couldn't account for \textit{any}
contemplation of facts already known to us---and thus, for the main activity
and one of the main subjects of philosophy itself!

We can now loosely state what Hintikka \cite{hintikka} called the
\textit{problem of logical omniscience}:

\begin{quotation}
\noindent Can we give some formal account of \textquotedblleft
knowledge\textquotedblright\ able to accommodate people learning new things
without leaving their armchairs?
\end{quotation}

Of course, one vacuous \textquotedblleft solution\textquotedblright\ would be
to declare that your knowledge\ is simply a list of all the true
sentences\footnote{If we don't require the sentences to be \textit{true}, then
presumably we're talking about \textit{belief} rather than \textit{knowledge}%
.} that you \textquotedblleft know\textquotedblright---and that, if the list
happens not to be closed under logical deductions, so be it! \ But this
\textquotedblleft solution\textquotedblright\ is no help at all at explaining
\textit{how} or \textit{why} you know things. \ Can't we do better?

Intuitively, we want to say that your \textquotedblleft
knowledge\textquotedblright\ consists of various non-logical facts
(\textquotedblleft grass is green\textquotedblright), together with
\textit{some} simple consequences of those facts (\textquotedblleft grass is
not pink\textquotedblright), but not necessarily \textit{all} the
consequences, and certainly not all consequences that involve difficult
mathematical reasoning. \ Unfortunately, as soon as we try to formalize this
idea, we run into problems.

The most obvious problem is the lack of a sharp boundary between the facts you
know right away, and those you \textquotedblleft could\textquotedblright%
\ know, but only after significant thought. \ (Recall the discussion of
\textquotedblleft known primes\textquotedblright\ from Section \ref{INTEGERS}%
.) \ A related problem is the lack of a sharp boundary between the facts you
know \textquotedblleft only if asked about them,\textquotedblright\ and those
you know even if you're \textit{not} asked. \ Interestingly, these two
boundaries seem to cut across each other. \ For example, while you've probably
already encountered the fact that $91$ is composite, it might take you some
time to remember it; while you've probably \textit{never} encountered the fact
that $83190$ is composite, once asked you can probably assent to it immediately.

But as discussed by Stalnaker \cite{stalnaker}, there's a third problem that
seems much more serious than either of the two above. \ Namely, you might
\textquotedblleft know\textquotedblright\ a particular fact if asked about it
one way, but not if asked in a different way! \ To illustrate this, Stalnaker
uses an example that we can recognize immediately from the discussion of the
$\mathsf{P}$ versus $\mathsf{NP}$\ problem in Section \ref{ENTSCH}. \ If I
asked you whether\ $43\times37=1591$, you could probably answer easily (e.g.,
by using $\left(  40+3\right)  \left(  40-3\right)  =40^{2}-3^{2}$). \ On the
other hand, if I instead asked you what the prime factors of $1591$ were, you
probably \textit{couldn't} answer so easily.

\begin{quotation}
\noindent But the answers to the two questions have the same content, even on
a very fine-grained notion of content. \ Suppose that we fix the threshold of
accessibility so that the information that $43$ and $37$ are the prime factors
of $1591$ is accessible in response to the second question, but not accessible
in response to the first. \ Do you know what the prime factors of $1591$ are
or not? ...\ Our problem is that we are not just trying to say what an agent
would know upon being asked certain questions; rather, we are trying to use
the facts about an agent's question answering capacities in order to get at
what the agent knows, even if the questions are not asked. \cite[p.
253]{stalnaker}
\end{quotation}

To add another example: does a typical four-year-old child \textquotedblleft
know\textquotedblright\ that addition of reals is commutative? \ Certainly not
if we asked her in those words---and if we tried to \textit{explain} the
words, she probably wouldn't understand us. \ Yet if we showed her a stack of
books, and asked her whether she could make the stack higher by shuffling the
books, she probably wouldn't make a mistake that involved imagining addition
was non-commutative. \ In that sense, we might say she already
\textquotedblleft implicitly\textquotedblright\ knows what her math classes
will later make explicit.

In my view, these and other examples strongly suggest that only a small part
of what we mean by \textquotedblleft knowledge\textquotedblright\ is knowledge
about the truth or falsehood of individual propositions. \ And crucially, this
remains so even if we restrict our attention to \textquotedblleft purely
verbalizable\textquotedblright\textit{ }knowledge---indeed, \textit{knowledge
used for answering factual questions}---and not (say) knowledge of how to ride
a bike or swing a golf club, or knowledge of a person or a place.\footnote{For
\textquotedblleft knowing\textquotedblright\ a person suggests having actually
met the person, while \textquotedblleft knowing\textquotedblright\ a place
suggests having visited the place. \ Interestingly, in Hebrew, one uses a
completely different verb for \textquotedblleft know\textquotedblright\ in the
sense of \textquotedblleft being familiar with\textquotedblright%
\ (\textit{makir}) than for \textquotedblleft know\textquotedblright\ in the
intellectual sense (\textit{yodeya}).} \ Many everyday uses of the word
\textquotedblleft know\textquotedblright\ support this idea:

\begin{quotation}
\noindent Do you know calculus?

\noindent Do you know Spanish?

\noindent Do you know the rules of bridge?
\end{quotation}

Each of the above questions could be interpreted as asking: \textit{do you
possess an internal algorithm, by which you can answer a large (and
possibly-unbounded) set of questions of some form? \ }While this is rarely
made explicit, the examples of this section and of Section \ref{INTEGERS}%
\ suggest adding the proviso: \textit{\ldots answer in a reasonable amount of
time?}

But suppose we accept that \textquotedblleft knowing how\textquotedblright%
\ (or \textquotedblleft knowing a good algorithm for\textquotedblright) is a
more fundamental concept than \textquotedblleft knowing
that.\textquotedblright\ \ How does that help us \textit{at all} in solving
the logical omniscience problem? \ You might worry that we're right back where
we started. \ After all, if we try to give a formal account of
\textquotedblleft knowing how,\textquotedblright\ then just like in the case
of \textquotedblleft knowing that,\textquotedblright\ it will be tempting to
write down axioms like the following:

\begin{quotation}
\noindent If you know how to compute $f\left(  x\right)  $ and $g\left(
x\right)  $ efficiently, then you also know how to compute $f\left(  x\right)
+g\left(  x\right)  $ efficiently.
\end{quotation}

\noindent Naturally, we'll then want to take the logical closure of those
axioms. \ But then, before we know it, won't we have conjured into our
imaginations a computationally-omniscient superbeing, who could efficiently
compute anything at all?

\subsection{The Cobham Axioms\label{COBHAM}}

Happily, the above worry turns out to be unfounded. \ We \textit{can} write
down reasonable axioms for \textquotedblleft knowing how to compute
efficiently,\textquotedblright\ and then \textit{go ahead and take the closure
of those axioms}, without getting the unwanted consequence of computational
omniscience. \ Explaining this point will involve a digression into an old and
fascinating corner of complexity theory---one that probably holds independent
interest for philosophers.

As is well-known, in the 1930s Church and Kleene proposed definitions of the
\textquotedblleft computable functions\textquotedblright\ that turned out to
be precisely equivalent to Turing's definition, but that differed from
Turing's in making no explicit mention of machines. \ Rather than analyzing
the \textit{process} of computation, the Church-Kleene approach was simply to
list \textit{axioms} that the computable functions of natural numbers
$f:\mathbb{N}\rightarrow\mathbb{N}$\ ought to satisfy---for example,
\textquotedblleft if $f\left(  x\right)  $ and $g\left(  x\right)  $\ are both
computable, then so is $f\left(  g\left(  x\right)  \right)  $%
\textquotedblright---and then to define \textquotedblleft
the\textquotedblright\ computable functions as the smallest set satisfying
those axioms.

In 1965, Alan Cobham \cite{cobham} asked whether the same could be done for
the \textit{efficiently} or \textit{feasibly} computable functions. \ As an
answer, he offered axioms that precisely characterize what today we call
$\mathsf{FP}$, or Function Polynomial-Time (though Cobham called it
$\mathcal{L}$). \ The class $\mathsf{FP}$\ consists\ of all functions of
natural numbers $f:\mathbb{N}\rightarrow\mathbb{N}$ that are computable in
polynomial time by a deterministic Turing machine. \ Note that $\mathsf{FP}%
$\ is \textquotedblleft morally\textquotedblright\ the same as the class
$\mathsf{P}$\ (Polynomial-Time) defined in Section \ref{ENTSCH}: they differ
only in that $\mathsf{P}$\ is a class of \textit{decision} problems (or
equivalently, functions $f:\mathbb{N}\rightarrow\left\{  0,1\right\}  $),
whereas $\mathsf{FP}$\ is a class of functions with integer range.

What was noteworthy about Cobham's characterization of polynomial time was
that it didn't involve \textit{any} explicit mention of either computing
devices or bounds on their running time. \ Let me now list a version of
Cobham's axioms, adapted from Arora, Impagliazzo, and Vazirani \cite{aiv}.
\ Each of the axioms talks about which functions of natural numbers
$f:\mathbb{N}\rightarrow\mathbb{N}$\ are \textquotedblleft efficiently
computable.\textquotedblright

\begin{enumerate}
\item[(1)] Every constant function $f$\ is efficiently computable, as is every
function\ which is nonzero only finitely often.

\item[(2)] \textbf{Pairing:} If $f\left(  x\right)  $ and $g\left(  x\right)
$ are efficiently computable, then so is $\left\langle f\left(  x\right)
,g\left(  x\right)  \right\rangle $, where $\left\langle ,\right\rangle $\ is
some standard pairing function for the natural numbers.

\item[(3)] \textbf{Composition:} If $f\left(  x\right)  $ and $g\left(
x\right)  $ are efficiently computable, then so is $f\left(  g\left(
x\right)  \right)  $.

\item[(4)] \textbf{Grab Bag:} The following functions are all efficiently computable:

\begin{itemize}
\item the arithmetic functions $x+y$ and $x\times y$

\item $\left\vert x\right\vert =\left\lfloor \log_{2}x\right\rfloor +1$ (the
number of bits in $x$'s binary representation)

\item the projection functions $\Pi_{1}\left(  \left\langle x,y\right\rangle
\right)  =x$\ and $\Pi_{2}\left(  \left\langle x,y\right\rangle \right)  =y$

\item $\operatorname*{bit}\left(  \left\langle x,i\right\rangle \right)
$\ (the $i^{th}$\ bit of $x$'s binary representation, or $0$ if $i>\left\vert
x\right\vert $)

\item $\operatorname*{diff}\left(  \left\langle x,i\right\rangle \right)  $
(the number obtained from $x$\ by flipping its $i^{th}$ bit)

\item $2^{\left\vert x\right\vert ^{2}}$ (called the \textquotedblleft smash
function\textquotedblright)
\end{itemize}

\item[(5)] \textbf{Bounded Recursion:} Suppose $f\left(  x\right)  $\ is
efficiently computable, and $\left\vert f\left(  x\right)  \right\vert
\leq\left\vert x\right\vert $ for all $x\in\mathbb{N}$. \ Then the function
$g\left(  \left\langle x,k\right\rangle \right)  $, defined by%
\[
g\left(  \left\langle x,k\right\rangle \right)  =\left\{
\begin{array}
[c]{cc}%
f\left(  g\left(  \left\langle x,\left\lfloor k/2\right\rfloor \right\rangle
\right)  \right)  & \text{if }k>1\\
x & \text{if }k=1
\end{array}
\right.  ,
\]
is also efficiently computable.
\end{enumerate}

A few comments about the Cobham axioms might be helpful. \ First, the axiom
that \textquotedblleft does most of the work\textquotedblright\ is (5).
\ Intuitively, given any natural number $k\in\mathbb{N}$ that we can generate
starting from the original input $x\in\mathbb{N}$, the Bounded Recursion axiom
lets us set up a \textquotedblleft computational process\textquotedblright%
\ that runs for $\log_{2}k$\ steps. \ Second, the role of the
\textquotedblleft smash function,\textquotedblright\ $2^{\left\vert
x\right\vert ^{2}}$, is to let us map $n$-bit integers to $n^{2}$-bit integers
to $n^{4}$-bit integers and so on, and thereby (in combination with the
Bounded Recursion axiom) set up computational processes that run for arbitrary
\textit{polynomial} numbers of steps. \ Third, although addition and
multiplication are included as \textquotedblleft efficiently computable
functions,\textquotedblright\ it is crucial that exponentiation is
\textit{not} included. \ Indeed, if $x$\ and $y$\ are $n$-bit integers, then
$x^{y}$\ might require exponentially many bits just to write down.

The basic result is then the following:

\begin{theorem}
[\cite{cobham,rose}]\label{cobhamthm}The class $\mathsf{FP}$, of functions
$f:\mathbb{N}\rightarrow\mathbb{N}$\ computable in polynomial time by a
deterministic Turing machine, satisfies axioms (1)-(5), and is the smallest
class that does so.
\end{theorem}

To prove Theorem \ref{cobhamthm}, one needs to do two things, neither of them
difficult: first, show that any function $f$\ that can be defined using the
Cobham axioms can also be computed in polynomial time; and second, show that
the Cobham axioms are enough to simulate any polynomial-time Turing machine.

One drawback of the Cobham axioms is that they seem to \textquotedblleft sneak
in the concept of polynomial-time through the back door\textquotedblright%
---both through the \textquotedblleft smash function,\textquotedblright\ and
through the arbitrary-looking condition $\left\vert f\left(  x\right)
\right\vert \leq\left\vert x\right\vert $\ in axiom (5). \ In the 1990s,
however, Leivant \cite{leivant} and Bellantoni and Cook \cite{bellantonicook}
both gave more \textquotedblleft elegant\textquotedblright\ logical
characterizations of $\mathsf{FP}$\ that avoid this problem. \ So for example,
Leivant showed that a function $f$\ belongs to $\mathsf{FP}$, if and only if
$f$ is computed by a program that can be proved correct in second-order logic
with comprehension restricted to positive quantifier-free formulas. \ Results
like these provide further evidence---if any was needed---that polynomial-time
computability is an extremely natural notion: a \textquotedblleft wide target
in conceptual space\textquotedblright\ that one hits even while aiming in
purely logical directions.

Over the past few decades, the idea of defining complexity classes such as
$\mathsf{P}$ and $\mathsf{NP}$\ in \textquotedblleft logical,
machine-free\textquotedblright\ ways has given rise to an entire field called
\textit{descriptive complexity theory}, which has deep connections with finite
model theory. \ While further discussion of descriptive complexity theory
would take us too far afield, see the book of Immerman \cite{immerman}\ for
the definitive introduction, or Fagin \cite{fagin}\ for a survey.

\subsection{Omniscience Versus Infinity\label{OMNIINF}}

Returning to our original topic, how exactly do axiomatic theories such as
Cobham's (or Church's and Kleene's, for that matter) escape the problem of
omniscience? \ One straightforward answer is that, unlike the set of true
sentences in some formal language, which is only \textit{countably} infinite,
the set of functions $f:\mathbb{N}\rightarrow\mathbb{N}$\ is
\textit{uncountably} infinite. \ And therefore, even if we define the
\textquotedblleft efficiently-computable\textquotedblright\ functions
$f:\mathbb{N}\rightarrow\mathbb{N}$\ by taking a countably-infinite logical
closure, we are sure to miss \textit{some} functions $f$\ (in fact, almost all
of them!).

The observation above suggests a general strategy to tame the logical
omniscience problem. \ Namely, we could refuse to define an agent's
\textquotedblleft knowledge\textquotedblright\ in terms of which individual
questions she can quickly answer, and insist on speaking instead about which
infinite \textit{families} of questions she can quickly answer. \ In slogan
form, we want to \textquotedblleft fight omniscience with
infinity.\textquotedblright

Let's see how, by taking this route, we can give semi-plausible answers to the
puzzles about knowledge discussed earlier in this section. \ First, the reason
why you can \textquotedblleft know\textquotedblright\ that $1591=43\times37$,
but at the same time \textit{not} \textquotedblleft know\textquotedblright%
\ the prime factors of $1591$, is that, when we speak about knowing the
answers to these questions, we really mean knowing \textit{how} to answer
them. \ And as we saw, there need not be any contradiction in knowing a fast
multiplication algorithm but \textit{not} a fast factoring algorithm, even if
we model your knowledge about algorithms as deductively closed. \ To put it
another way, by embedding the two questions

\begin{quotation}
\noindent Q1 =~\textquotedblleft Is $1591=43\times37$?\textquotedblright

\noindent Q2 = \textquotedblleft What are the prime factors of $1591$%
?\textquotedblright
\end{quotation}

\noindent into \textit{infinite families of related questions}, we can break
the symmetry between the knowledge entailed in answering them.

Similarly, we could think of a child as possessing an internal algorithm
which, given any statement of the form $x+y=y+x$\ (for specific $x$\ and
$y$\ values), immediately outputs \textit{true}, without even examining $x$
and $y$. \ However, the child does not yet have the ability to process
\textit{quantified} statements, such as \textquotedblleft$\forall
x,y\in\mathbb{R}~~x+y=y+x$.\textquotedblright\ \ In that sense, she still
lacks the explicit knowledge that addition is commutative.

Although the \textquotedblleft cure\textquotedblright\ for logical
omniscience\ sketched above solves some puzzles, not surprisingly it raises
many puzzles of its own. \ So let me end this section by discussing three
major objections to the \textquotedblleft infinity cure.\textquotedblright

The first objection is that we've simply pushed the problem of logical
omniscience\ somewhere else. \ For suppose an agent \textquotedblleft
knows\textquotedblright\ how to compute every function in some restricted
class such as $\mathsf{FP}$. \ Then how can we ever make sense of the agent
\textit{learning a new algorithm?} \ One natural response is that,\ even if
you have the \textquotedblleft latent ability\textquotedblright\ to compute a
function $f\in\mathsf{FP}$, you might not \textit{know} that you have the
ability---either because you don't know a suitable algorithm, or because you
\textit{do} know an algorithm, but don't know that it's an algorithm for $f$.
\ Of course, if we wanted to pursue things to the bottom, we'd next need to
tell a story about \textit{knowledge of algorithms}, and how logical
omniscience is avoided there. \ However, I claim that this represents
progress! \ For notice that, even without such a story, we can already explain
\textit{some} failures of logical omniscience. \ For example, the reason why
you don't know the factors of a large number might \textit{not} be your
ignorance of a fast factoring method, but rather that no such method exists.

The second objection is that, when I advocated focusing on infinite families
of questions rather than single questions in isolation, I never specified
\textit{which} infinite families. \ The difficulty is that the same question
could be generalized in wildly different ways. \ As an example, consider the question

\begin{quotation}
\noindent Q = \textquotedblleft Is $432,150$\ composite?\textquotedblright
\end{quotation}

\noindent Q is an instance of a computational problem that humans find very
hard: \textquotedblleft given a large integer $N$, is $N$
composite?\textquotedblright\ \ However, Q is \textit{also} an instance of a
computational problem that humans find very easy: \textquotedblleft given a
large integer $N$ \textit{ending in} $0$, is $N$ composite?\textquotedblright%
\ \ And indeed, we'd expect a person to know the answer to Q \textit{if} she
noticed that $432,150$ ends in $0$, but not otherwise. \ To me, what this
example demonstrates is that, \textit{if} we want to discuss an agent's
knowledge in terms of individual questions such as Q, then the relevant issue
will be whether there \textit{exists} a generalization G of Q, such that the
agent knows a fast algorithm for answering questions of type G, and
\textit{also} recognizes that Q is of type G.

The third objection is just the standard one about the relationship between
asymptotic complexity and finite statements. \ For example, if we model an
agent's knowledge using the Cobham axioms, then we can indeed explain why the
agent doesn't know how to play perfect chess on an $n\times n$\ board, for
\textit{arbitrary} values of $n$.\footnote{For chess on an $n\times n$\ board
is known to be $\mathsf{EXP}$-complete, and it is also known that
$\mathsf{P}\neq\mathsf{EXP}$. \ See Section \ref{TIME}, and particularly
footnote \ref{chessnote}, for more details.} \ But on a standard $8\times
8$\ board, playing perfect chess would \textquotedblleft
merely\textquotedblright\ require (say) $\sim10^{60}$ computational steps,
which is a constant, and therefore certainly polynomial! \ So strictly on the
basis of the Cobham axioms, what explanation could we possibly offer for why a
rational agent, who knew the rules of $8\times8$ chess, didn't also know how
to play it optimally? \ While this objection might sound devastating, it's
important to understand that it's no different from the usual objection
leveled against complexity-theoretic arguments, and can be given the usual
response. \ Namely: asymptotic statements are \textit{always} vulnerable to
being rendered irrelevant, if the constant factors turned out to be
ridiculous. \ However, experience has shown that, for whatever reasons, that
happens rarely enough that one can usually take asymptotic behavior as
\textquotedblleft having explanatory force until proven
otherwise.\textquotedblright\ \ (Section \ref{CONC} will say more about the
explanatory force of asymptotic claims, as a problem requiring philosophical analysis.)

\subsection{Summary}

Because of the difficulties pointed out in Section \ref{OMNIINF}, my own view
is that computational complexity theory has not yet come close to
\textquotedblleft solving\textquotedblright\ the logical omniscience problem,
in the sense of giving a satisfying formal account of knowledge that also
avoids making absurd predictions. \ I have no idea whether such an account is
even possible.\footnote{Compare the pessimism expressed by Paul Graham
\cite{graham} about knowledge representation more generally:
\par
\begin{quotation}
\noindent In practice formal logic is not much use, because despite some
progress in the last 150 years we're still only able to formalize a small
percentage of statements. \ We may never do that much better, for the same
reason 1980s-style \textquotedblleft knowledge
representation\textquotedblright\ could never have worked; many statements may
have no representation more concise than a huge, analog brain state.
\end{quotation}
} \ However, what I've tried to show in this section is that complexity theory
provides a well-defined \textquotedblleft limiting case\textquotedblright%
\ where the logical omniscience problem \textit{is} solvable, about as well as
one could hope it to be. \ The limiting case is where the size of the
questions grows without bound, and the solution there is given by the Cobham
axioms: \textquotedblleft axioms of knowing how\textquotedblright\ whose
logical closure one \textit{can} take without thereby inviting omniscience.

In other words, when we contemplate the omniscience problem, I claim that
we're in a situation similar to one often faced in physics---where we might be
at a loss to understand some phenomenon (say, gravitational entropy),
\textit{except} in limiting cases such as black holes. \ In epistemology just
like in physics, the limiting cases that we \textit{do} more-or-less
understand offer an obvious starting point for those wishing to tackle the
general case.

\section{Computationalism and Waterfalls\label{WATERFALLS}}

Over the past two decades, a certain argument about computation---which I'll
call the \textit{waterfall argument}---has been widely discussed by
philosophers of mind.\footnote{See Putnam \cite[appendix]{putnam}\ and Searle
\cite{searle:redis}\ for two instantiations of the argument (though the formal
details of either will not concern us here).} \ Like Searle's famous Chinese
Room argument \cite{searle}, the waterfall argument seeks to show that
computations are \textquotedblleft inherently syntactic,\textquotedblright%
\ and can never be \textquotedblleft about\textquotedblright\ anything---and
that for this reason, the doctrine of \textquotedblleft
computationalism\textquotedblright\ is false.\footnote{\textquotedblleft
Computationalism\textquotedblright\ refers to the view that the mind is
literally\ a computer, and that thought is literally a type of computation.}
\ But unlike the Chinese Room, the waterfall argument supplements the bare
appeal to intuition by a further claim: namely, that the \textquotedblleft
meaning\textquotedblright\ of a computation, to whatever extent it has one, is
always \textit{relative to some external observer}.

More concretely, consider a waterfall (though any other physical system with a
large enough state space would do as well). \ Here I do not mean a waterfall
that was specially engineered to perform computations, but \textit{really} a
naturally-occurring waterfall: say, Niagara Falls. \ Being governed by laws of
physics, the waterfall implements some mapping $f$\ from\ a set\ of possible
initial states to a set of possible final states. \ If we accept that the laws
of physics are \textit{reversible}, then $f$ must also be injective. \ Now
suppose we restrict attention to some finite subset $S$ of possible initial
states, with $\left\vert S\right\vert =n$. \ Then $f$ is just a one-to-one
mapping from $S$\ to some output set $T=f\left(  S\right)  $\ with $\left\vert
T\right\vert =n$. \ The \textquotedblleft crucial
observation\textquotedblright\ is now this: given \textit{any} permutation
$\sigma$\ from the set of integers $\left\{  1,\ldots,n\right\}  $\ to itself,
there is some way to label the elements of $S$ and $T$ by integers in
$\left\{  1,\ldots,n\right\}  $, such that we can interpret $f$\ as
implementing $\sigma$. \ For example, if we let\ $S=\left\{  s_{1}%
,\ldots,s_{n}\right\}  $\ and $f\left(  s_{i}\right)  =t_{i}$, then it
suffices to label the initial state $s_{i}$\ by $i$ and the final state
$t_{i}$\ by $\sigma\left(  i\right)  $. \ But the permutation $\sigma$\ could
have any \textquotedblleft semantics\textquotedblright\ we like: it might
represent a program for playing chess, or factoring integers, or simulating a
different waterfall. \ Therefore \textquotedblleft mere
computation\textquotedblright\ cannot give rise to semantic meaning. \ Here is
how Searle \cite[p. 57]{searle:redis} expresses the conclusion:

\begin{quotation}
\noindent If we are consistent in adopting the Turing test or some other
\textquotedblleft objective\textquotedblright\ criterion for intelligent
behavior, then the answer to such questions as \textquotedblleft Can
unintelligent bits of matter produce intelligent behavior?\textquotedblright%
\ and even, \textquotedblleft How exactly do they do it\textquotedblright\ are
ludicrously obvious. \ Any thermostat, pocket calculator, or waterfall
produces \textquotedblleft intelligent behavior,\textquotedblright\ and we
know in each case how it works. \ Certain artifacts are designed to behave as
if they were intelligent, and since everything follows laws of nature, then
everything will have some description under which it behaves as if it were
intelligent. \ But this sense of \textquotedblleft intelligent
behavior\textquotedblright\ is of no psychological relevance at all.
\end{quotation}

The waterfall argument has been criticized on numerous grounds: see Haugeland
\cite{haugeland}, Block \cite{block}, and especially Chalmers \cite{chalmers}
(who parodied the argument by proving that a cake recipe, being merely
syntactic, can never give rise to the semantic attribute of crumbliness). \ To
my mind, though, perhaps the easiest way to demolish the waterfall argument is
through\ computational complexity considerations.

Indeed, suppose we actually wanted to use a waterfall to help us calculate
chess moves. \ How would we do that? \ In complexity terms, what we want is a
\textit{reduction} from the chess problem to the waterfall-simulation problem.
\ That is, we want an efficient algorithm that somehow \textit{encodes} a
chess position $P$ into an initial state $s_{P}\in S$ of the waterfall, in
such a way that a good move from $P$ can be read out efficiently from the
waterfall's corresponding final state, $f\left(  s_{P}\right)  \in
T$.\footnote{Technically, this describes a restricted class of reductions,
called \textit{nonadaptive} reductions. \ An \textit{adaptive} reduction from
chess to waterfalls might solve a chess problem by some procedure that
involves initializing a waterfall and observing its final state, then using
the results of that aquatic computation to initialize a \textit{second}
waterfall and observe \textit{its} final state, and so on for some polynomial
number of repetitions.} \ But \textit{what would such an algorithm look like?}
\ We cannot say for sure---certainly not without detailed knowledge about
$f$\ (i.e., the physics of waterfalls), as well as the means by which the $S$
and $T$\ elements are encoded as binary strings. \ But for \textit{any}
reasonable choice, it seems overwhelmingly likely that any reduction algorithm
would just \textit{solve the chess problem itself}, without using the
waterfall in an essential way at all! \ A bit more precisely, I conjecture
that, given any chess-playing algorithm $A$\ that accesses a \textquotedblleft
waterfall oracle\textquotedblright\ $W$, there is an equally-good
chess-playing algorithm $A^{\prime}$, with similar time and space
requirements, that does \textit{not} access $W$. \ If this conjecture holds,
then it gives us a perfectly observer-independent way to formalize our
intuition that the \textquotedblleft semantics\textquotedblright\ of
waterfalls have nothing to do with chess.\footnote{The perceptive reader might
suspect that we smuggled our conclusion into the assumption that the waterfall
states $s_{P}\in S$\ and $f\left(  s_{P}\right)  \in T$\ were encoded as
binary strings in a \textquotedblleft reasonable\textquotedblright\ way (and
not, for example, in a way that encodes the solution to the chess problem).
\ But a crucial lesson of complexity theory is that, when we discuss
\textquotedblleft computational problems,\textquotedblright\ we
\textit{always} make an implicit commitment about the input and output
encodings anyway! \ So for example, if positive integers were given as input
via their prime factorizations, then the factoring problem would be trivial
(just apply the identity function). \ But who cares? \ If, in mathematically
defining the waterfall-simulation problem, we required input and output
encodings that entailed solving chess problems, then it would no longer be
reasonable to call our problem (solely) a \textquotedblleft
waterfall-simulation problem\textquotedblright\ at all.}

\subsection{\textquotedblleft Reductions\textquotedblright\ That Do All The
Work\label{REDUCTIONS}}

Interestingly, the issue of \textquotedblleft trivial\textquotedblright\ or
\textquotedblleft degenerate\textquotedblright\ reductions also arises
\textit{within} complexity theory, so it might be instructive to see how it is
handled there. \ Recall from Section \ref{ENTSCH}\ that a problem is
$\mathsf{NP}$\textit{-complete} if, loosely speaking, it is \textquotedblleft
maximally hard among all $\mathsf{NP}$ problems\textquotedblright%
\ ($\mathsf{NP}$ being the class of problems for which solutions can be
checked in polynomial time). \ More formally, we say that $L$ is $\mathsf{NP}$-complete\ if

\begin{enumerate}
\item[(i)] $L\in\mathsf{NP}$, and

\item[(ii)] given any \textit{other} $\mathsf{NP}$\ problem $L^{\prime}$,
there exists a polynomial-time algorithm to solve $L^{\prime}$ using access to
an oracle that solves $L$. \ (Or more succinctly, $L^{\prime}\in\mathsf{P}%
^{L}$, where $\mathsf{P}^{L}$\ denotes the complexity class $\mathsf{P}%
$\ augmented by an $L$-oracle.)
\end{enumerate}

The concept of $\mathsf{NP}$-completeness had incredible explanatory power: it
showed that \textit{thousands} of seemingly-unrelated problems from physics,
biology, industrial optimization, mathematical logic, and other fields were
all \textit{identical} from the standpoint of polynomial-time computation, and
that not one of these problems had an efficient solution unless $\mathsf{P}%
=\mathsf{NP}$. \ Thus, it was natural for theoretical computer scientists to
want to define an analogous concept of $\mathsf{P}$\textit{-completeness}.
\ In other words: among all the problems that \textit{are} solvable in
polynomial time, which ones are \textquotedblleft maximally
hard\textquotedblright?

But how should $\mathsf{P}$-completeness even be defined? \ To see the
difficulty, suppose that, by analogy with $\mathsf{NP}$-completeness, we say
that $L$\ is $\mathsf{P}$-complete if

\begin{enumerate}
\item[(i)] $L\in\mathsf{P}$ and

\item[(ii)] $L^{\prime}\in\mathsf{P}^{L}$\ for every $L^{\prime}\in\mathsf{P}$.
\end{enumerate}

Then it is easy to see that the second condition is vacuous: \textit{every}
$\mathsf{P}$\ problem is $\mathsf{P}$-complete! \ For in \textquotedblleft
reducing\textquotedblright\ $L^{\prime}$\ to $L$, a polynomial-time algorithm
can always just ignore the $L$-oracle\ and solve $L^{\prime}$\ by itself, much
like our hypothetical chess program that ignored its waterfall oracle.
\ Because of this, condition (ii) must be replaced by a stronger condition;
one popular choice is

\begin{enumerate}
\item[(ii')] $L^{\prime}\in\mathsf{LOGSPACE}^{L}$\ for every $L^{\prime}%
\in\mathsf{P}$.
\end{enumerate}

Here $\mathsf{LOGSPACE}$\ means, informally, the class of problems solvable by
a deterministic Turing machine with a read/write memory\ consisting of only
$\log n$\ bits, given an input of size $n$.\footnote{Note that a
$\mathsf{LOGSPACE}$\ machine does not even have enough memory to store its
input string! \ For this reason, we think of the input string as being
provided on a special \textit{read-only} tape.} \ It's not hard to show that
$\mathsf{LOGSPACE}\subseteq\mathsf{P}$, and this containment is strongly
believed to be strict (though just like with $\mathsf{P}\neq\mathsf{NP}$,
there is no proof yet). \ The key point is that, if we want a
\textit{non-vacuous} notion of completeness, then the reducing complexity
class needs to be \textit{weaker} (either provably or conjecturally) than the
class being reduced to. \ In fact complexity classes even smaller than
$\mathsf{LOGSPACE}$\ almost always suffice in practice.

In my view, there is an important lesson here for debates about
computationalism. \ Suppose we want to claim, for example, that a computation
that plays chess is \textquotedblleft equivalent\textquotedblright\ to some
other computation that simulates a waterfall. \ Then our claim is only
non-vacuous if it's possible to \textit{exhibit} the equivalence (i.e., give
the reductions) within a model of computation that isn't \textit{itself}
powerful enough to solve the chess or waterfall problems.

\section{PAC-Learning and the Problem of Induction\label{PAC}}

Centuries ago, David Hume \cite{hume} famously pointed out that learning from
the past (and, by extension, science) seems logically impossible. \ For
example, if we sample $500$ ravens and every one of them is black, why does
that give us \textit{any} grounds---even probabilistic grounds---for expecting
the $501^{st}$\ raven to be black also? \ Any modern answer to this question
would probably refer to \textit{Occam's razor}, the principle that simpler
hypotheses consistent with the data are more likely to be correct. \ So for
example, the hypothesis that all ravens are black is \textquotedblleft
simpler\textquotedblright\ than the hypothesis that most ravens are green or
purple, and that only the $500$ we happened to see were black. \ Intuitively,
it seems Occam's razor \textit{must} be part of the solution to Hume's
problem; the difficulty is that such a response leads to questions of its own:

\begin{enumerate}
\item[(1)] What do we mean by \textquotedblleft simpler\textquotedblright?

\item[(2)] \textit{Why} are simple explanations likely to be correct? \ Or,
less ambitiously: what properties must reality have for Occam's Razor to
\textquotedblleft work\textquotedblright?

\item[(3)] How much data must we collect before we can find a
\textquotedblleft simple hypothesis\textquotedblright\ that will probably
predict future data? \ How do we go about finding such a hypothesis?
\end{enumerate}

In my view, the theory of \textit{PAC (Probabilistically Approximately
Correct) Learning}, initiated by Leslie Valiant \cite{valiant:pac}\ in 1984,
has made large enough advances on all of these questions that it deserves to
be studied by anyone interested in induction.\footnote{See Kearns and Vazirani
\cite{kvaz} for an excellent introduction to PAC-learning, and de Wolf
\cite{dewolf:masters} for previous work applying PAC-learning to philosophy
and linguistics: specifically, to fleshing out Chomsky's \textquotedblleft
poverty of the stimulus\textquotedblright\ argument. \ De Wolf also discusses
several formalizations of Occam's Razor other than the one based on
PAC-learning.} \ In this theory,\ we consider an idealized \textquotedblleft
learner,\textquotedblright\ who is presented with points $x_{1},\ldots,x_{m}%
$\ drawn randomly\ from some large set $\mathcal{S}$, together with the
\textquotedblleft classifications\textquotedblright\ $f\left(  x_{1}\right)
,\ldots,f\left(  x_{m}\right)  $ of those points. \ The learner's goal is to
infer the function $f$, well enough to be able to predict $f\left(  x\right)
$\ for \textit{most} future points $x\in\mathcal{S}$. \ As an example, the
learner might be a bank, $\mathcal{S}$\ might be a set of people (represented
by their credit histories), and $f\left(  x\right)  $\ might represent whether
or not person $x$\ will default on a loan.

For simplicity, we often assume that $\mathcal{S}$\ is a set of binary
strings, and that the function $f$ maps each $x\in\mathcal{S}$\ to a single
bit, $f\left(  x\right)  \in\left\{  0,1\right\}  $. \ Both assumptions can be
removed without significantly changing the theory. \ The important assumptions
are the following:

\begin{enumerate}
\item[(1)] Each of the sample points $x_{1},\ldots,x_{m}$\ is drawn
\textit{independently} from some (possibly-unknown) \textquotedblleft sample
distribution\textquotedblright\ $\mathcal{D}$ over $\mathcal{S}$.
\ Furthermore, the future points $x$ on which the learner will need to predict
$f\left(  x\right)  $ are drawn from the same distribution.

\item[(2)] The function $f$\ belongs to a known \textquotedblleft hypothesis
class\textquotedblright\ $\mathcal{H}$. \ This $\mathcal{H}$ represents
\textquotedblleft the set of possibilities the learner is willing to
entertain\textquotedblright\ (and is typically much smaller than the set of
all $2^{\left\vert \mathcal{S}\right\vert }$\ possible functions from
$\mathcal{S}$\ to $\left\{  0,1\right\}  $).
\end{enumerate}

Under these assumptions, we have the following central result.

\begin{theorem}
[Valiant \cite{valiant:pac}]\label{valthm}Consider a finite hypothesis class
$\mathcal{H}$, a Boolean function $f:\mathcal{S}\rightarrow\left\{
0,1\right\}  $ in $\mathcal{H}$, and a sample distribution $\mathcal{D}$ over
$\mathcal{S}$, as well as an error rate $\varepsilon>0$ and failure
probability $\delta>0$ that the learner is willing to tolerate. \ Call a
hypothesis $h:\mathcal{S}\rightarrow\left\{  0,1\right\}  $\ \textquotedblleft
good\textquotedblright\ if%
\[
\Pr_{x\sim\mathcal{D}}\left[  h\left(  x\right)  =f\left(  x\right)  \right]
\geq1-\varepsilon.
\]
Also, call sample points $x_{1},\ldots,x_{m}$\ \textquotedblleft
reliable\textquotedblright\ if any hypothesis $h\in\mathcal{H}$\ that
satisfies $h\left(  x_{i}\right)  =f\left(  x_{i}\right)  $\ for all
$i\in\left\{  1,\ldots,m\right\}  $ is good. \ Then%
\[
m=\frac{1}{\varepsilon}\ln\frac{\left\vert \mathcal{H}\right\vert }{\delta}%
\]
sample points $x_{1},\ldots,x_{m}$\ drawn independently from $\mathcal{D}%
$\ will be reliable with probability at least $1-\delta$.
\end{theorem}

Intuitively, Theorem \ref{valthm} says that the behavior of $f$ on a small
number of randomly-chosen points \textit{probably} determines its behavior on
\textit{most} of the remaining points. \ In other words, if, by some
unspecified means, the learner manages to find any hypothesis $h\in
\mathcal{H}$\ that makes correct predictions on all its past data points
$x_{1},\ldots,x_{m}$, then provided $m$ is large enough (and as it happens,
$m$ doesn't need to be very large), the learner can be statistically confident
that $h$ will also make the correct predictions on most future points.

The part of Theorem \ref{valthm}\ that bears the unmistakable imprint of
complexity theory is the bound on sample size, $m\geq\frac{1}{\varepsilon}%
\ln\frac{\left\vert \mathcal{H}\right\vert }{\delta}$. \ This bound has three
notable implications. \ First, even if the class $\mathcal{H}$\ contains
exponentially many hypotheses (say, $2^{n}$), one can still learn an arbitrary
function $f\in\mathcal{H}$\ using a \textit{linear} amount of sample data,
since $m$ grows only logarithmically with $\left\vert \mathcal{H}\right\vert
$: in other words, like the number of bits needed to \textit{write down} an
individual hypothesis. \ Second, one can make the probability that the
hypothesis $h$\ will fail to generalize \textit{exponentially small} (say,
$\delta=2^{-n}$), at the cost of increasing the sample size $m$ by only a
linear factor. \ Third, assuming the hypothesis \textit{does} generalize, its
error rate $\varepsilon$\ decreases inversely with $m$. \ It is not hard to
show that each of these dependencies is tight, so that for example, if we
demand either $\varepsilon=0$\ or $\delta=0$\ then no finite $m$ suffices.
\ This is the origin of the name \textquotedblleft
PAC-learning\textquotedblright: the most one can hope for is to output a
hypothesis that is \textquotedblleft probably, approximately\textquotedblright\ correct.

The proof of Theorem \ref{valthm} is easy: consider any hypothesis
$h\in\mathcal{H}$ that is \textit{bad}, meaning that%
\[
\Pr_{x\sim\mathcal{D}}\left[  h\left(  x\right)  =f\left(  x\right)  \right]
<1-\varepsilon.
\]
Then by the independence assumption,%
\[
\Pr_{x_{1},\ldots,x_{m}\sim\mathcal{D}}\left[  h\left(  x_{1}\right)
=f\left(  x_{1}\right)  \wedge\cdots\wedge h\left(  x_{m}\right)  =f\left(
x_{m}\right)  \right]  <\left(  1-\varepsilon\right)  ^{m}.
\]
Now, the number of bad hypotheses is no more than the total number of
hypotheses, $\left\vert \mathcal{H}\right\vert $. \ So by the union bound, the
probability that there \textit{exists} a bad hypothesis that agrees with
$f$\ on all of $x_{1},\ldots,x_{m}$\ can be at most $\left\vert \mathcal{H}%
\right\vert \cdot\left(  1-\varepsilon\right)  ^{m}$. \ Therefore $\delta
\leq\left\vert \mathcal{H}\right\vert \cdot\left(  1-\varepsilon\right)  ^{m}%
$, and all that remains is to solve for $m$.

The relevance of Theorem \ref{valthm}\ to Hume's problem of induction is that
the theorem describes a nontrivial class of situations where induction is
\textit{guaranteed to work} with high probability.\ \ Theorem \ref{valthm}
also illuminates the role of Occam's Razor in induction. \ In order to learn
using a \textquotedblleft reasonable\textquotedblright\ number of sample
points $m$, the hypothesis class $\mathcal{H}$\ must have a sufficiently small
cardinality. \ But that is equivalent to saying that every hypothesis
$h\in\mathcal{H}$\ must have a \textit{succinct description}---since the
number of bits needed to specify an arbitrary hypothesis $h\in\mathcal{H}$ is
simply $\left\lceil \log_{2}\left\vert \mathcal{H}\right\vert \right\rceil $.
\ If the number of bits needed to specify a hypothesis is too large, then
$\mathcal{H}$\ will always be vulnerable to the problem of
\textit{overfitting}: some hypotheses $h\in\mathcal{H}$\ surviving contact
with the sample data just by chance.

As pointed out to me by Agust\'{\i}n Rayo, there are several possible
interpretations of Occam's Razor that have nothing to do with descriptive
complexity: for example, we might want our hypotheses to be \textquotedblleft
simple\textquotedblright\ in terms of their ontological or ideological
commitments. \ However, to whatever extent we interpret Occam's Razor as
saying that \textit{shorter} or \textit{lower-complexity} hypotheses are
preferable, Theorem \ref{valthm} comes closer than one might have thought
possible to a mathematical justification for why the Razor works.

Many philosophers might be familiar with alternative formal approaches to
Occam's Razor. \ For example, within a Bayesian framework, one can choose a
prior over all possible hypotheses that gives greater weight to
\textquotedblleft simpler\textquotedblright\ hypotheses (where simplicity is
measured, for example, by the length of the shortest program that computes the
predictions). \ However, while the PAC-learning and Bayesian approaches are
related, the PAC approach has the advantage of requiring only a
\textit{qualitative} decision about which hypotheses one wants to consider,
rather than a quantitative prior over hypotheses. \ Given the hypothesis class
$\mathcal{H}$, one can then seek learning methods that work for \textit{any}
$f\in\mathcal{H}$. \ (On the other hand, the PAC approach requires an
assumption about the probability distribution over \textit{observations},
while the Bayesian approach does not.)

\subsection{Drawbacks of the Basic PAC Model\label{DRAWBACKS}}

I'd now like to discuss three drawbacks of Theorem \ref{valthm}, since I think
the drawbacks illuminate philosophical aspects of induction as well as the
advantages do.

The first drawback is that Theorem \ref{valthm} works only for \textit{finite}
hypothesis classes. \ In science, however, hypotheses often involve continuous
parameters, of which there is an uncountable infinity. \ Of course, one could
solve this problem by simply discretizing the parameters, but then the number
of hypotheses (and therefore the relevance of Theorem \ref{valthm}) would
depend on how fine the discretization was. \ Fortunately, we can avoid such
difficulties by realizing that \textit{the learner only cares about the
\textquotedblleft differences\textquotedblright\ between two hypotheses
insofar as they lead to different predictions.} \ This leads to the
fundamental notion of \textit{VC-dimension} (after its originators, Vapnik and
Chervonenkis \cite{vc}).

\begin{definition}
[VC-dimension]A hypothesis class $\mathcal{H}$\ shatters the sample points
$\left\{  x_{1},\ldots,x_{k}\right\}  \subseteq\mathcal{S}$\ if for all
$2^{k}$\ possible settings of $h\left(  x_{1}\right)  ,\ldots,h\left(
x_{k}\right)  $, there exists a hypothesis $h\in\mathcal{H}$ compatible with
those settings. \ Then $\operatorname*{VCdim}\left(  \mathcal{H}\right)  $,
the VC-dimension of $\mathcal{H}$, is the largest $k$ for which there exists a
subset $\left\{  x_{1},\ldots,x_{k}\right\}  \subseteq\mathcal{S}$\ that
$\mathcal{H}$\ shatters (or if no finite maximum exists, then
$\operatorname*{VCdim}\left(  \mathcal{H}\right)  =\infty$).
\end{definition}

Clearly any finite hypothesis class has finite VC-dimension: indeed,
$\operatorname*{VCdim}\left(  \mathcal{H}\right)  \leq\log_{2}\left\vert
\mathcal{H}\right\vert $. \ However, even an infinite hypothesis class can
have finite VC-dimension if it is \textquotedblleft sufficiently
simple.\textquotedblright\ \ For example, let $\mathcal{H}$ be the class of
all functions $h_{a,b}:\mathbb{R}\rightarrow\left\{  0,1\right\}  $\ of the
form%
\[
h_{a,b}\left(  x\right)  =\left\{
\begin{array}
[c]{cc}%
1 & \text{if }a\leq x\leq b\\
0 & \text{otherwise.}%
\end{array}
\right.
\]
Then it is easy to check that $\operatorname*{VCdim}\left(  \mathcal{H}%
\right)  =2$.

With the notion of VC-dimension in hand, we can state a powerful (and
harder-to-prove!) generalization of Theorem \ref{valthm},\ due to Blumer et
al.\ \cite{behw}.

\begin{theorem}
[Blumer et al.\ \cite{behw}]\label{vcthm}For some universal constant $K>0$,
the bound on $m$ in Theorem \ref{valthm}\ can be replaced by%
\[
m=\frac{K\operatorname*{VCdim}\left(  \mathcal{H}\right)  }{\varepsilon}%
\ln\frac{1}{\delta\varepsilon},
\]
with the theorem now holding for any hypothesis class $\mathcal{H}$, finite or infinite.
\end{theorem}

If $\mathcal{H}$ has infinite VC-dimension, then it is easy to construct a
probability distribution $\mathcal{D}$\ over sample points such that
\textit{no finite number }$m$\textit{\ of samples from }$D$\textit{\ suffices
to PAC-learn a function} $f\in\mathcal{H}$: one really is in the unfortunate
situation described by Hume, of having no grounds at all for predicting that
the next raven will be black. \ In some sense, then, Theorem \ref{vcthm}\ is
telling us that finite VC-dimension is a necessary and sufficient condition
for scientific induction to be possible. \ Once again, Theorem \ref{vcthm}%
\ also has an interpretation in terms of Occam's Razor, with the smallness of
the VC-dimension now playing the role of simplicity.

The second drawback of Theorem \ref{valthm}\ is that it gives us no clues
about how to \textit{find} a hypothesis $h\in\mathcal{H}$\ consistent with the
sample data. \ All it says is that, \textit{if} we find such an $h$, then $h$
will probably be close to the truth. \ This illustrates that, even in the
simple setup envisioned by PAC-learning, induction \textit{cannot} be merely a
matter of seeing enough data and then \textquotedblleft
generalizing\textquotedblright\ from it, because immense computations might be
needed to \textit{find} a suitable generalization! \ Indeed, following the
work of Kearns and Valiant \cite{kv}, we now know that many natural learning
problems---as an example, inferring the rules of a regular or context-free
language from random examples of grammatical and ungrammatical sentences---are
computationally intractable in an extremely strong sense:

\begin{quotation}
\noindent\textit{Any polynomial-time algorithm for finding a hypothesis consistent with
the data would imply a polynomial-time algorithm for breaking widely-used
cryptosystems such as RSA!}\footnote{In the setting of \textquotedblleft
proper learning\textquotedblright---where the learner needs to output a
hypothesis in some specified format---it is even known that many natural
PAC-learning problems are $\mathsf{NP}$-complete (see Pitt and Valiant
\cite{pittvaliant} for example). \ But in the \textquotedblleft
improper\textquotedblright\ setting---where the learner can describe its
hypothesis using any polynomial-time algorithm---it is only known how to show
that PAC-learning problems are hard under cryptographic assumptions, and there
seem to be inherent reasons for this (see Applebaum, Barak, and Xiao
\cite{abx}).}
\end{quotation}

The appearance of \textit{cryptography} in the above statement is far from
accidental. \ In a sense that can be made precise, learning and cryptography
are \textquotedblleft dual\textquotedblright\ problems: a learner wants to
find patterns in data, while a cryptographer wants to generate data whose
patterns are \textit{hard} to find. \ More concretely, one of the basic
primitives in cryptography is called a \textit{pseudorandom function family}.
\ This is a family of efficiently-computable Boolean functions $f_{s}:\left\{
0,1\right\}  ^{n}\rightarrow\left\{  0,1\right\}  $, parameterized by a short
random \textquotedblleft seed\textquotedblright\ $s$, that are
\textit{virtually} \textit{indistinguishable\ from random functions} by a
polynomial-time algorithm. \ Here, we imagine that the would-be distinguishing
algorithm can query the function $f_{s}$ on various points $x$, and also that
it \textit{knows} the mapping from $s$\ to $f_{s}$, and so is ignorant only of
the seed $s$\ itself. \ There is strong evidence in cryptography that
pseudorandom function families exist: indeed, Goldreich, Goldwasser, and
Micali \cite{ggm} showed how to construct one starting from any pseudorandom
\textit{generator} (the latter was mentioned in Section \ref{WONT}).

Now, given a pseudorandom function family $\left\{  f_{s}\right\}  $, imagine
a PAC-learner whose hypothesis class $\mathcal{H}$\ consists of $f_{s}$\ for
all possible seeds\ $s$. \ The learner is provided some randomly-chosen sample
points $x_{1},\ldots,x_{m}\in\left\{  0,1\right\}  ^{n}$, together with the
values of $f_{s}$\ on those points: $f_{s}\left(  x_{1}\right)  ,\ldots
,f_{s}\left(  x_{m}\right)  $. \ Given this \textquotedblleft training
data,\textquotedblright\ the learner's goal is to figure out how to compute
$f_{s}$\ for itself---and thereby predict the values of $f_{s}\left(
x\right)  $\ on new points $x$, points \textit{not} in the training sample.
\ Unfortunately, it's easy to see that \textit{if} the learner could do that,
then it would thereby distinguish $f_{s}$\ from a truly random function---and
thereby contradict our starting assumption that $\left\{  f_{s}\right\}
$\ was pseudorandom! \ Our conclusion is that, \textit{if} the basic
assumptions of modern cryptography hold (and in particular, if there exist
pseudorandom generators), then there must be situations where learning is
impossible purely because of computational complexity (and not because of
insufficient data).

The third drawback of Theorem \ref{valthm}\ is the assumption that the
distribution $\mathcal{D}$\ from which the learner is tested is the same as
the distribution from which the sample points were drawn. \ To me, this is the
most serious drawback, since it tells us that\ PAC-learning models the
\textquotedblleft learning\textquotedblright\ performed by an undergraduate
cramming for an exam by solving last year's problems, or an employer using a
regression model to identify the characteristics of successful hires, or a
cryptanalyst breaking a code from a collection of plaintexts and ciphertexts.
\ It does not, however, model the \textquotedblleft learning\textquotedblright%
\ of an Einstein or a Szilard, making predictions about phenomena that are
different in kind from anything yet observed. \ As David Deutsch stresses in
his recent book \textit{The Beginning of Infinity} \cite{deutsch:infinity},
the goal of science is not merely to summarize observations, and thereby let
us make predictions about similar observations. \ Rather, the goal is to
discover explanations with \textquotedblleft reach,\textquotedblright\ meaning
the ability to predict what would happen even in novel or hypothetical
situations, like the Sun suddenly disappearing or a quantum computer being
built. \ In my view, developing a compelling mathematical model of
\textit{explanatory} learning---a model that \textquotedblleft is to
explanation as the PAC model is to prediction\textquotedblright---is an
outstanding open problem.\footnote{Important progress toward this goal
includes the work of Angluin \cite{angluin}\ on learning finite automata from
queries and counterexamples, and that of Angluin et al.\ \cite{aacw}\ on
learning a circuit by injecting values. \ Both papers study natural learning
models that generalize the PAC model by allowing \textquotedblleft controlled
scientific experiments,\textquotedblright\ whose results confirm or refute a
hypothesis and thereby provide guidance about which experiments to do next.}

\subsection{Computational Complexity, Bleen, and Grue\label{GRUE}}

In 1955, Nelson Goodman \cite{goodman}\ proposed what he called the
\textquotedblleft new riddle of induction,\textquotedblright\ which survives
the Occam's Razor answer to Hume's original induction problem. \ In Goodman's
riddle, we are asked to consider the hypothesis \textquotedblleft All emeralds
are green.\textquotedblright\ \ The question is, why do we favor \textit{that}
hypothesis over the following alternative, which is equally compatible with
all our evidence of green emeralds?

\begin{quotation}
\noindent\textquotedblleft All emeralds are green before January 1, 2030, and
then blue afterwards.\textquotedblright
\end{quotation}

The obvious answer is that the second hypothesis adds superfluous
complications, and is therefore disfavored by Occam's Razor. \ To that,
Goodman replies that the definitions of \textquotedblleft
simple\textquotedblright\ and \textquotedblleft complicated\textquotedblright%
\ depend on our language. \ In particular, suppose we had no words for green
or blue, but we did have a word \textit{grue}, meaning \textquotedblleft green
before January 1, 2030, and blue afterwards,\textquotedblright\ and a word
\textit{bleen}, meaning \textquotedblleft blue before January 1, 2030, and
green afterwards.\textquotedblright\ \ In that case, we could only express the
hypothesis \textquotedblleft All emeralds are green\textquotedblright\ by saying

\begin{quotation}
\noindent\textquotedblleft All emeralds are grue before January 1, 2030, and
then bleen afterwards.\textquotedblright
\end{quotation}

\noindent---a manifestly more complicated hypothesis than the simple
\textquotedblleft All emeralds are grue\textquotedblright!

I confess that, when I contemplate the grue riddle, I can't help but recall
the joke about the Anti-Inductivists, who, when asked why they continue to
believe that the future \textit{won't} resemble the past, when that false
belief has brought their civilization nothing but poverty and misery, reply,
\textquotedblleft because anti-induction has never worked
before!\textquotedblright\ \ Yes, if we artificially define our primitive
concepts \textquotedblleft against the grain of the world,\textquotedblright%
\ then we shouldn't be surprised if the world's actual behavior becomes more
cumbersome to describe, or if we make wrong predictions. \ It would be as if
we were using a programming language that had no built-in function for
multiplication, but only for $F\left(  x,y\right)  :=17x-y-x^{2}+2xy$. \ In
that case, a normal person's first instinct would be either to switch
programming languages, or else to \textit{define}\ multiplication in terms of
$F$, and forget about $F$\ from that point onward!\footnote{Suppose that our
programming language provides only multiplication by constants, addition, and
the function $F\left(  x,y\right)  :=ax^{2}+bxy+cy^{2}+dx+ey+f$. \ We can
assume without loss of generality that $d=e=f=0$. \ Then provided
$ax^{2}+bxy+cy^{2}$\ factors into two independent linear terms, $px+qy$\ and
$rx+sy$, we can express the product $xy$\ as%
\[
\frac{F\left(  sx-qy,-rx+py\right)  }{\left(  ps-qr\right)  ^{2}}.
\]
} \ Now, there \textit{is} a genuine philosophical problem here: why
\textit{do} grue, bleen, and $F\left(  x,y\right)  $\ go \textquotedblleft
against the grain of the world,\textquotedblright\ whereas green, blue, and
multiplication go with the grain? \ But to me, that problem (like Wigner's
puzzlement over \textquotedblleft the unreasonable effectiveness of
mathematics in natural sciences\textquotedblright\ \cite{wigner}) is more
about the world itself than about human concepts, so we shouldn't expect any
purely linguistic analysis to resolve it.

What about computational complexity, then? \ In my view, while computational
complexity doesn't solve the grue riddle, it does contribute a useful insight.
\ Namely, that when we talk about the simplicity or complexity of hypotheses,
we should distinguish two issues:

\begin{enumerate}
\item[(a)] The \textit{asymptotic scaling} of the hypothesis size, as the
\textquotedblleft size\textquotedblright\ $n$ of our learning problem goes to infinity.

\item[(b)] The constant-factor overheads.
\end{enumerate}

In terms of the basic PAC model in Section \ref{PAC}, we can imagine a
\textquotedblleft hidden parameter\textquotedblright\ $n$, which measures the
number of bits needed to specify an individual point in the set $\mathcal{S}%
=\mathcal{S}_{n}$. \ (Other ways to measure the \textquotedblleft
size\textquotedblright\ of a learning problem would also work, but this way is
particularly convenient.) \ For convenience, we can identify $\mathcal{S}_{n}$
with the set$\ \left\{  0,1\right\}  ^{n}$ of $n$-bit strings, so that
$n=\log_{2}\left\vert \mathcal{S}_{n}\right\vert $. \ We then need to
consider, not just a \textit{single} hypothesis class, but an infinite
\textit{family} of hypothesis classes $\mathcal{H}=\left\{  \mathcal{H}%
_{1},\mathcal{H}_{2},\mathcal{H}_{3},\ldots\right\}  $, one for each positive
integer $n$. \ Here $\mathcal{H}_{n}$\ consists of hypothesis functions\ $h$%
\ that map $\mathcal{S}_{n}=\left\{  0,1\right\}  ^{n}$\ to $\left\{
0,1\right\}  $.

Now let $L$ be a \textit{language} for specifying hypotheses in $\mathcal{H}$:
in other words, a mapping from (some subset of) binary strings $y\in\left\{
0,1\right\}  ^{\ast}$\ to $\mathcal{H}$. \ Also, given a hypothesis
$h\in\mathcal{H}$, let%
\[
\kappa_{L}\left(  h\right)  :=\min\left\{  \left\vert y\right\vert :L\left(
y\right)  =h\right\}
\]
be the length of the \textit{shortest} description of $h$ in the language $L$.
\ (Here $\left\vert y\right\vert $\ just means the number of bits in $y$.)
\ Finally, let%
\[
\kappa_{L}\left(  n\right)  :=\max\left\{  \kappa_{L}\left(  h\right)
:h\in\mathcal{H}_{n}\right\}
\]
be the number of bits needed to specify an \textit{arbitrary} hypothesis in
$\mathcal{H}_{n}$ using the language $L$. \ Clearly $\kappa_{L}\left(
n\right)  \geq\left\lceil \log_{2}\left\vert \mathcal{H}_{n}\right\vert
\right\rceil $, with equality if and only if $L$\ is \textquotedblleft
optimal\textquotedblright\ (that is, if it represents each hypothesis
$h\in\mathcal{H}_{n}$\ using as few bits as possible). \ The question that
concerns us is how quickly $\kappa_{L}\left(  n\right)  $\ grows as a function
of $n$, for various choices of language $L$.

What does any of this have to do with the grue riddle? \ Well, we can think of
the details of $L$\ (its syntax, vocabulary, etc.) as affecting the
\textquotedblleft lower-order\textquotedblright\ behavior of the function
$\kappa_{L}\left(  n\right)  $. \ So for example, suppose we are unlucky
enough that $L$ contains the words \textit{grue} and \textit{bleen}, but not
\textit{blue} and \textit{green}. \ That might increase $\kappa_{L}\left(
n\right)  $\ by a factor of ten or so---since now, every time we want to
mention \textquotedblleft green\textquotedblright\ when specifying our
hypothesis $h$, we instead need a wordy circumlocution like \textquotedblleft
grue before January 1, 2030, and then bleen afterwards,\textquotedblright\ and
similarly for blue.\footnote{Though note that, if the language \thinspace$L$
is expressive enough to allow this, we can simply define green and blue in
terms of bleen and grue \textit{once}, then refer back to those definitions
whenever needed! \ In that case, taking bleen and grue (rather than green and
blue) to be the primitive concepts would increase $\kappa_{L}\left(  n\right)
$\ by only an \textit{additive} constant, rather than a multiplicative
constant.
\par
The above fact is related to a fundamental result from the theory of
Kolmogorov complexity (see Li and Vit\'{a}nyi \cite{livitanyi}\ for example).
\ Namely, if $P$ and $Q$ are any two Turing-universal programming languages,
and if $K_{P}\left(  x\right)  $\ and $K_{Q}\left(  x\right)  $\ are the
lengths of the shortest programs in $P$ and $Q$ respectively that output a
given string $x\in\left\{  0,1\right\}  ^{\ast}$, then there exists a
universal \textquotedblleft translation constant\textquotedblright\ $c_{PQ}$,
such that $\left\vert K_{P}\left(  x\right)  -K_{Q}\left(  x\right)
\right\vert \leq c_{PQ}$ for \textit{every} $x$. \ This $c_{PQ}$\ is just the
number of bits needed to write a $P$-interpreter for $Q$-programs or vice
versa.} \ However, a crucial lesson of complexity theory is that the
\textquotedblleft higher-order\textquotedblright\ behavior of $\kappa
_{L}\left(  n\right)  $---for example, whether it grows polynomially or
exponentially with $n$---is almost completely unaffected by the details of
$L$! \ The reason is that, if two languages $L_{1}$\ and $L_{2}$\ differ only
in their \textquotedblleft low-level details,\textquotedblright\ then
\textit{translating} a hypothesis from $L_{1}$\ to $L_{2}$\ or vice versa will
increase the description length by no more than a polynomial factor. \ Indeed,
as in our grue example, there is usually a \textquotedblleft universal
translation constant\textquotedblright\ $c$ such that $\kappa_{L_{1}}\left(
h\right)  \leq c\kappa_{L_{2}}\left(  h\right)  $\ or even $\kappa_{L_{1}%
}\left(  h\right)  \leq\kappa_{L_{2}}\left(  h\right)  +c$\ for \textit{every}
hypothesis $h\in\mathcal{H}$.

The one exception to the above rule is if the languages $L_{1}$ and $L_{2}$
have different\textit{ expressive powers}. \ For example, maybe $L_{1}$ only
allows nesting expressions to depth two, while $L_{2}$ allows nesting to
arbitrary depths; or $L_{1}$ only allows propositional connectives, while
$L_{2}$ also allows first-order quantifiers. \ In those cases, $\kappa_{L_{1}%
}\left(  h\right)  $ could indeed be much greater than $\kappa_{L_{2}}\left(
h\right)  $\ for some hypotheses $h$, possibly even exponentially greater
($\kappa_{L_{1}}\left(  h\right)  \approx2^{\kappa_{L_{2}}\left(  h\right)  }%
$). \ A rough analogy would be this: suppose you hadn't learned what
differential equations were, and had no idea how to solve them even
approximately or numerically. \ In that case, Newtonian mechanics might seem
just as complicated to you as the Ptolemaic theory with epicycles, if not
\textit{more} complicated! \ For the only way you could make predictions with
Newtonian mechanics would be using a huge table of \textquotedblleft
precomputed\textquotedblright\ differential equation solutions---and
\textit{to you}, that table would seem just as unwieldy and inelegant as a
table of epicycles. \ But notice that in this case, your perception would be
the result, not of some arbitrary choice of vocabulary, but of an
\textit{objective} gap in your mathematical expressive powers.

To summarize, our choice of vocabulary---for example, whether we take
green/blue or bleen/grue as primitive concepts---could indeed matter\ if we
want to use Occam's Razor to predict the future color of emeralds. \ But I
think that complexity theory justifies us in treating grue as a
\textquotedblleft small-$n$ effect\textquotedblright:\ something that becomes
less and less important in the asymptotic limit of more and more complicated
learning problems.

\section{Quantum Computing\label{QC}}

\textit{Quantum computing} is a proposal for using quantum mechanics to solve
certain computational problems much faster than we know how to solve them
today.\footnote{The authoritative reference for quantum computing is the book
of Nielsen and Chuang \cite{nc}. \ For gentler introductions, try Mermin
\cite{mermin,mermin:book}\ or the survey articles of Aharonov
\cite{aharonov:review}, Fortnow \cite{fortnow:qc}, or Watrous
\cite{watrous:survey}. \ For a general discussion of polynomial-time
computation and the laws of physics (including speculative models beyond
quantum computation), see my survey article \textquotedblleft$\mathsf{NP}%
$-complete Problems and Physical Reality\textquotedblright\ \cite{aar:np}.}
\ To do so, one would need to build a new type of computer, capable of
exploiting the quantum effects of superposition and interference. \ Building
such a computer---one large enough to solve interesting problems---remains an
enormous challenge for physics and engineering, due to the fragility of
quantum states and the need to isolate them from their external environment.

In the meantime, though, theoretical computer scientists have extensively
studied what we could and couldn't do with a quantum computer if we had one.
\ For certain problems, remarkable quantum algorithms are known to solve them
in polynomial time, even though the best-known classical algorithms require
exponential time. \ Most famously, in 1994 Peter Shor \cite{shor}\ gave a
polynomial-time quantum algorithm for factoring integers, and as a byproduct,
breaking most of the cryptographic codes used on the Internet today. \ Besides
the practical implications, Shor's algorithm also provided a key piece of
evidence that switching from classical to quantum computers would enlarge the
class of problems solvable in polynomial-time. \ For theoretical computer
scientists, this had a profound lesson: if we want to know the limits of
efficient computation, we may need to \textquotedblleft leave our
armchairs\textquotedblright\ and incorporate actual facts about physics (at a
minimum, the truth or falsehood of quantum mechanics!).\footnote{By contrast,
if we only want to know what is \textit{computable} in the physical universe,
with no efficiency requirement, then it remains entirely consistent with
current knowledge that Church and Turing gave the correct answer\ in the
1930s---and that they did so without incorporating any physics beyond what is
\textquotedblleft accessible to intuition.\textquotedblright}

Whether or not scalable quantum computers are built anytime soon, my own
(biased) view is that quantum computing represents one of the great scientific
advances of our time. \ But here I want to ask a different question: does
quantum computing have any implications for \textit{philosophy}---and
specifically, for the interpretation of quantum mechanics?

From one perspective, the answer seems like an obvious \textquotedblleft
no.\textquotedblright\ \ Whatever else it is, quantum computing is
\textquotedblleft merely\textquotedblright\ an application of quantum
mechanics, as that theory has existed in physics textbooks for 80 years.
\ Indeed,\ \textit{if} you accept that quantum mechanics (as currently
understood) is true, then presumably you should also accept the possibility of
quantum computers, and make the same predictions about their operation as
everyone else. \ Whether you describe the \textquotedblleft
reality\textquotedblright\ behind quantum processes via the Many-Worlds
Interpretation, Bohmian mechanics, or some other view (or, following Bohr's
Copenhagen Interpretation, refuse to discuss the \textquotedblleft
reality\textquotedblright\ at all), seems irrelevant.

From a different perspective, though, a scalable quantum computer would
\textit{test }quantum mechanics\textit{ }in an extremely novel regime---and
for that reason, it could indeed raise new philosophical issues. \ The
\textquotedblleft regime\textquotedblright\ quantum computers would test is
characterized not by an energy scale or a temperature, but by computational
complexity. \ One of the most striking facts about quantum mechanics is that,
to represent the state of $n$ entangled particles, one needs a vector of size
\textit{exponential} in $n$. \ For example, to specify the state\ of a
thousand spin-$1/2$ particles, one needs $2^{1000}$\ complex numbers called
\textquotedblleft amplitudes,\textquotedblright\ one for every possible
outcome of measuring the spins in the $\left\{  \operatorname*{up}%
,\operatorname*{down}\right\}  $ basis. \ The quantum state, denoted
$\left\vert \psi\right\rangle $,\ is then a linear combination or
\textquotedblleft superposition\textquotedblright\ of the possible outcomes,
with each outcome $\left\vert x\right\rangle $\ weighted by its amplitude
$\alpha_{x}$:%
\[
\left\vert \psi\right\rangle =\sum_{x\in\left\{  \operatorname*{up}%
,\operatorname*{down}\right\}  ^{1000}}\alpha_{x}\left\vert x\right\rangle .
\]
Given $\left\vert \psi\right\rangle $, one can calculate the probability
$p_{x}$\ that any particular outcome $\left\vert x\right\rangle $ will be
observed, via the rule $p_{x}=\left\vert \alpha_{x}\right\vert ^{2}%
$.\footnote{This means, in particular, that the amplitudes satisfy the
normalization condition $\sum_{x}\left\vert \alpha_{x}\right\vert ^{2}=1$.}

Now, there are only about $10^{80}$\ atoms in the visible universe, which is a
much smaller number than $2^{1000}$. \ So assuming quantum mechanics is true,
it seems Nature has to invest \textit{staggering} amounts of \textquotedblleft
computational effort\textquotedblright\ to keep track of small collections of
particles---certainly more than anything classical physics
requires!\footnote{One might object that even in the classical world, if we
simply don't \textit{know} the value of (say) an $n$-bit string, then we
\textit{also} describe our ignorance using exponentially-many numbers: namely,
the \textit{probability} $p_{x}$ of each possible string $x\in\left\{
0,1\right\}  ^{n}$! \ And indeed, there is an extremely close connection
between quantum mechanics and classical probability theory; I often describe
quantum mechanics as just \textquotedblleft probability theory with complex
numbers instead of nonnegative reals.\textquotedblright\ \ However, a crucial
difference is that we can always describe a classical string $x$\ as
\textquotedblleft really\textquotedblright\ having a definite value; the
vector of $2^{n}$\ probabilities $p_{x}$\ is then just a mental representation
of our own ignorance. \ With a quantum state, we do not have the same luxury,
because of the phenomenon of \textit{interference} between positive and
negative amplitudes.}$^{,}$\footnote{One might also object that, even in
classical physics, it takes \textit{infinitely} many bits to record the state
of even a single particle, if its position and momentum can be arbitrary real
numbers. \ And indeed, Copeland \cite{copeland}, Hogarth \cite{hogarth},
Siegelmann \cite{siegelmann}, and other writers have speculated that the
continuity of physical quantities might actually allow \textquotedblleft
hypercomputations\textquotedblright---including solving the halting problem in
a finite amount of time! \ From a modern perspective, though, quantum
mechanics and quantum gravity strongly suggest that \textit{the
\textquotedblleft continuity\textquotedblright\ of measurable quantities such
as positions and momenta is a theoretical artifact.} \ In other words, it
ought to suffice for simulation purposes to approximate these quantities to
some finite precision, probably related to the Planck scale of $10^{-33}%
$\ centimeters or $10^{-43}$\ seconds.
\par
But the exponentiality of quantum states is different, for at least two
reasons. \ Firstly, it doesn't lead to computational speedups that are nearly
as \textquotedblleft unreasonable\textquotedblright\ as the hypercomputing
speedups. \ Secondly, no one has any idea where the theory in question
(quantum mechanics) \textit{could} break down, in a manner consistent with
current experiments. \ In other words, there is no known \textquotedblleft
killer obstacle\textquotedblright\ for quantum computing, analogous to the
Planck scale for hypercomputing. \ See Aaronson \cite{aar:mlin}\ for further
discussion of this point, as well as a proposed complexity-theoretic framework
(called \textquotedblleft Sure/Shor separators\textquotedblright) with which
to study such obstacles.} \ In the early 1980s, Richard Feynman
\cite{feynman:qc} and others called attention to this point, noting that it
underlay something that had long been apparent in practice: the extraordinary
difficulty of simulating quantum mechanics using conventional computers. \ But
Feynman also raised the possibility of turning that difficulty around, by
building our computers out of quantum components. \ Such computers could
conceivably solve certain problems faster than conventional computers: if
nothing else, then at least the problem of simulating quantum mechanics!

Thus, quantum computing is interesting not just because of its applications,
but (even more, in my opinion) because \textit{it is the first technology that
would directly \textquotedblleft probe\textquotedblright\ the exponentiality
inherent in the quantum description of Nature.} \ One can make an analogy here
to the experiments in the 1980s that first convincingly violated the Bell
Inequality. \ Like quantum algorithms today, Bell's refutation of local
realism was \textquotedblleft merely\textquotedblright\ a mathematical
consequence of quantum mechanics.\ \ But that refutation (and the experiments
that it inspired) made conceptually-important \textit{aspects} of quantum
mechanics no longer possible to ignore---and for that reason, it changed the
philosophical landscape. \ It seems overwhelmingly likely to me that quantum
computing will do the same.

Indeed, we can extend the analogy further: just as there were
\textquotedblleft local realist diehards\textquotedblright\ who denied that
Bell Inequality violation would be possible (and tried to explain it away
after it was achieved), so today a vocal minority of computer scientists and
physicists (including Leonid Levin \cite{levin:qc}, Oded Goldreich
\cite{goldreich:qc}, and Gerard 't Hooft \cite{thooft})\ denies the
possibility of scalable quantum computers, even in principle. \ While they
admit that quantum mechanics has passed every experimental test for a century,
these skeptics are \textit{confident} that quantum mechanics will fail in the
regime tested by quantum computing---and that whatever new theory replaces it,
that theory will allow only classical computing.

As most quantum computing researchers are quick to point out in response, they
would be \textit{thrilled} if the attempt to build scalable quantum computers
led instead to a revision of quantum mechanics! \ Such an outcome would
probably constitute the largest revolution in physics since the 1920s, and
ultimately be much \textit{more} interesting than building a quantum computer.
\ Of course, it is also possible that scalable quantum computing will be given
up as too difficult for \textquotedblleft mundane\textquotedblright%
\ technological reasons, rather than fundamental physics reasons. \ But that
\textquotedblleft mundane\textquotedblright\ possibility is not what skeptics
such as Levin, Goldreich, and 't Hooft are talking about.

\subsection{Quantum Computing and the Many-Worlds Interpretation\label{MWI}}

But let's return to the original question: suppose the skeptics are wrong, and
it \textit{is} possible to build scalable quantum computers. \ Would that have
any relevance to the interpretation of quantum mechanics? \ The best-known
argument that the answer is \textquotedblleft yes\textquotedblright\ was made
by David Deutsch, a quantum computing pioneer\ and staunch defender of the
Many-Worlds Interpretation. \ To be precise, Deutsch thinks that quantum
mechanics \textit{straightforwardly} implies the existence of parallel
universes, and that it does so independently of quantum computing: on his
view, even the double-slit experiment can only be explained in terms of two
parallel universes interfering. \ However, Deutsch also thinks that quantum
computing adds emotional punch to the argument. \ Here is how he put it in his
1997 book \textit{The Fabric of Reality} \cite[p. 217]{deutsch}:

\begin{quotation}
\noindent Logically, the possibility of complex quantum computations adds
nothing to a case [for the Many-Worlds Interpretation] that is already
unanswerable. \ But it does add psychological impact. \ With Shor's algorithm,
the argument has been writ very large. \ To those who still cling to a
single-universe world-view, I issue this challenge: \textit{explain how Shor's
algorithm works.} \ I do not merely mean predict that it will work, which is
merely a matter of solving a few uncontroversial equations. \ I mean provide
an explanation. \ When Shor's algorithm has factorized a number, using
$10^{500}$\ or so times the computational resources that can be seen to be
present, where was the number factorized? \ There are only about $10^{80}%
$\ atoms in the entire visible universe, an utterly minuscule number compared
with $10^{500}$. \ So if the visible universe were the extent of physical
reality, physical reality would not even remotely contain the resources
required to factorize such a large number. \ Who did factorize it, then?
\ How, and where, was the computation performed?
\end{quotation}

There is plenty in the above paragraph for an enterprising philosopher to
mine. \ In particular, how \textit{should} a nonbeliever in Many-Worlds answer
Deutsch's challenge? \ In the rest of this section, I'll focus on two possible responses.

The first response is to deny that, if Shor's algorithm works as predicted,
that can only be explained by postulating \textquotedblleft vast computational
resources.\textquotedblright\ \ At the most obvious level, complexity
theorists have not yet ruled out the possibility of a fast \textit{classical}
factoring algorithm.\footnote{Indeed, one \textit{cannot} rule that
possibility out, without first proving $\mathsf{P}\neq\mathsf{NP}$! \ But even
if $\mathsf{P}\neq\mathsf{NP}$, a fast classical factoring algorithm might
\textit{still} exist, again because factoring is not thought to be
$\mathsf{NP}$-complete.} \ More generally, that quantum computers can solve
certain problems superpolynomially faster than classical computers is not a
theorem, but a (profound, plausible) \textit{conjecture}.\footnote{A formal
version of this conjecture is $\mathsf{BPP}\neq\mathsf{BQP}$, where
$\mathsf{BPP}$\ (Bounded-Error Probabilistic Polynomial-Time) and
$\mathsf{BQP}$\ (Bounded-Error Quantum Polynomial-Time)\ are the classes of
problems efficiently solvable by classical randomized algorithms and quantum
algorithms respectively. \ Bernstein and Vazirani \cite{bv}\ showed that
$\mathsf{P}\subseteq\mathsf{BPP}\subseteq\mathsf{BQP}\subseteq\mathsf{PSPACE}%
$, where $\mathsf{PSPACE}$\ is the class of problems solvable by a
deterministic Turing machine using a polynomial amount of \textit{memory} (but
possibly exponential time). \ For this reason, any proof of the $\mathsf{BPP}%
\neq\mathsf{BQP}$\ conjecture would immediately imply $\mathsf{P}%
\neq\mathsf{PSPACE}$ as well. \ The latter would be considered almost as great
a breakthrough as $\mathsf{P}\neq\mathsf{NP}$.}$^{,}$\footnote{Complicating
matters, there \textit{are} quantum algorithms that provably achieve
exponential speedups over any classical algorithm: one example is Simon's
algorithm \cite{simon}, an important predecessor of Shor's algorithm.
\ However, all such algorithms are formulated in the \textquotedblleft
black-box model\textquotedblright\ (see Beals et al.\ \cite{bbcmw}),\ where
the resource to be minimized is the number of queries that an algorithm makes
to a hypothetical black box. \ Because it is relatively easy to analyze, the
black-box model is a crucial source of insights about what \textit{might} be
true in the conventional Turing machine model. \ However, it is also known
that the black-box model sometimes misleads us about the \textquotedblleft
real\textquotedblright\ situation. \ As a famous example, the complexity
classes $\mathsf{IP}$\ and $\mathsf{PSPACE}$\ are equal \cite{shamir}, despite
the existence of a black box that separates them (see Fortnow
\cite{fortnow:rel}\ for discussion).
\par
Besides the black-box model, \textit{unconditional}\ exponential separations
between quantum and classical complexities are known in several other
restricted models, including communication complexity \cite{raz:cc}.} \ If the
conjecture failed, then the door would seem open to what we might call
\textquotedblleft polynomial-time hidden-variable theories\textquotedblright:
theories that reproduce the predictions of quantum mechanics without invoking
any computations outside $\mathsf{P}$.\footnote{Technically, if the
hidden-variable theory involved classical randomness, then it would correspond
more closely to the complexity class $\mathsf{BPP}$\ (Bounded-Error
Probabilistic Polynomial-Time). \ However, today there is strong evidence that
$\mathsf{P}=\mathsf{BPP}$ (see Impagliazzo and Wigderson \cite{iw}).} \ These
would be analogous to the \textit{local} hidden variable theories that
Einstein and others had hoped for, before Bell ruled such theories out.

A second response to Deutsch's challenge is that, even if we agree that Shor's
algorithm demonstrates the reality of vast \textit{computational resources} in
Nature, it is not obvious that we should think of those resources as
\textquotedblleft parallel universes.\textquotedblright\ \ Why not simply say
that there is \textit{one} universe, and that it is quantum-mechanical?
\ Doesn't the parallel-universes language reflect an ironic
\textit{parochialism}: a desire to impose a familiar science-fiction image on
a mathematical theory that is \textit{stranger} than fiction, that doesn't
match \textit{any} of our pre-quantum intuitions (including computational
intuitions) particularly well?

One can sharpen the point as follows: \textit{if} one took the
parallel-universes explanation of how a quantum computer works too seriously
(as many popular writers do!), then it would be natural to make further
inferences about quantum computing that are flat-out wrong. \ For example:

\begin{quotation}
\noindent\textit{\textquotedblleft Using only a thousand quantum bits (or
qubits), a quantum computer could store }$2^{1000}$\textit{\ classical
bits.\textquotedblright}
\end{quotation}

This is true only for a bizarre definition of the word \textquotedblleft
store\textquotedblright! \ The fundamental problem is that, when you measure a
quantum computer's state, you see only \textit{one} of the possible outcomes;
the rest disappear. \ Indeed, a celebrated result called \textit{Holevo's
Theorem} \cite{holevo}\ says that, using $n$ qubits, there is no way to store
more than $n$ classical bits so that the bits can be reliably retrieved later.
\ In other words: for at least one natural definition of \textquotedblleft
information-carrying capacity,\textquotedblright\ qubits have exactly the same
capacity as bits.

To take another example:

\begin{quotation}
\noindent\textit{\textquotedblleft Unlike a classical computer, which can only
factor numbers by trying the divisors one by one, a quantum computer could try
all possible divisors in parallel.\textquotedblright}
\end{quotation}

If quantum computers can harness vast numbers of parallel worlds, then the
above seems like a reasonable guess as to how Shor's algorithm works. \ But
\textit{it's not how it works at all}. \ Notice that, if Shor's algorithm
\textit{did} work that way, then it could be used not only for factoring
integers, but also for the much larger task of solving $\mathsf{NP}$-complete
problems in polynomial time. \ (As mentioned in footnote \ref{FACTOR}, the
factoring problem is strongly believed \textit{not} to be $\mathsf{NP}%
$-complete.) \ But contrary to a common misconception, quantum computers are
neither known nor believed to be able to solve $\mathsf{NP}$-complete problems
efficiently.\footnote{There is a remarkable quantum algorithm called
\textit{Grover's algorithm} \cite{grover}, which can search any space of
$2^{N}$\ possible solutions in only $\thicksim2^{N/2}$\ steps. \ However,
Grover's algorithm represents a \textit{quadratic} (square-root) improvement
over classical brute-force search, rather than an exponential improvement.
\ And without any further assumptions about the structure of the search space,
Grover's algorithm is optimal, as shown by Bennett et al.\ \cite{bbbv}.} \ As
usual, the fundamental problem is that measuring reveals just a single random
outcome $\left\vert x\right\rangle $. \ To get around that problem, and ensure
that the \textit{right} outcome is observed with high probability, a quantum
algorithm needs to generate an \textit{interference pattern}, in which the
computational paths leading to a given wrong outcome cancel each other out,
while the paths leading to a given right outcome reinforce each other. \ This
is a delicate requirement, and as far as anyone knows, it can only be achieved
for a few problems, most of which (like the factoring problem) have special
structure arising from algebra or number theory.\footnote{Those interested in
further details of how Shor's algorithm works, but still not ready for a
mathematical exposition, might want to try my popular essay \textquotedblleft
Shor, I'll Do It\textquotedblright\ \cite{aar:shor}.}

A Many-Worlder might retort: \textquotedblleft sure, I agree that quantum
computing involves harnessing the parallel universes in subtle and non-obvious
ways, but it's still \textit{harnessing parallel universes!}\textquotedblright%
\ \ But even here, there's a fascinating irony. \ Suppose we choose to think
of a quantum algorithm in terms of parallel universes.\ \ Then to put it
crudely, not only must many universes interfere to give a large final
amplitude to the right answer; they must also, by interfering, \textit{lose
their identities as parallel universes!} \ In other words, to whatever extent
a collection of universes is useful for quantum computation, to that extent it
is arguable whether we ought to call them \textquotedblleft parallel
universes\textquotedblright\ at all (as opposed to parts of one
exponentially-large, self-interfering, quantum-mechanical blob). \ Conversely,
to whatever extent the universes have unambiguously separate identities, to
that extent they're now \textquotedblleft decohered\textquotedblright\ and out
of causal contact with each other. Thus we can explain the outputs of any
future computations by invoking only one of the universes, and treating the
others as unrealized hypotheticals.

To clarify, I don't regard either of the above objections to Deutsch's
argument as decisive, and am unsure what I think about the matter. \ My
purpose, in setting out the objections, was simply to illustrate the potential
of quantum computing theory to inform debates about the Many-Worlds Interpretation.

\section{New Computational Notions of Proof\label{PROOF}}

Since the time of Euclid, there have been two main notions of mathematical proof:

\begin{enumerate}
\item[(1)] A \textquotedblleft proof\textquotedblright\ is a verbal
explanation that induces a sense of certainty (and ideally, understanding)
about the statement to be proved, in any human mathematician willing and able
to follow it.

\item[(2)] A \textquotedblleft proof\textquotedblright\ is a finite sequence
of symbols encoding syntactic deductions in some formal system, which start
with axioms and end with the statement to be proved.
\end{enumerate}

The tension between these two notions is a recurring theme in the philosophy
of mathematics. \ But theoretical computer science deals regularly with a
third notion of proof---one that seems to have received much less
philosophical analysis than either of the two above. \ This notion is the following:

\begin{enumerate}
\item[(3)] A \textquotedblleft proof\textquotedblright\ is any computational
process or protocol (real or imagined) that can terminate in a certain way if
and only if the statement to be proved is true.
\end{enumerate}

\subsection{Zero-Knowledge Proofs\label{ZKP}}

As an example of this third notion, consider \textit{zero-knowledge proofs},
introduced by Goldwasser, Micali, and Rackoff \cite{gmr}. \ Given two graphs
$G$\ and $H$, each with $n\approx10000$ vertices, suppose that an all-powerful
but untrustworthy wizard Merlin wishes to convince a skeptical king Arthur
that $G$ and $H$ are \textit{not} isomorphic. \ Of course, one way Merlin
could do this would be to list all $n!$\ graphs obtained by permuting the
vertices of $G$, then note that none of these equal $H$. \ However, such a
proof would clearly exhaust Arthur's patience (indeed, it could not even be
written down within the observable universe). \ Alternatively, Merlin could
point Arthur to some \textit{property} of $G$ and $H$ that differentiates
them: for example, maybe their adjacency matrices have different eigenvalue
spectra. \ Unfortunately, it is not yet proven that, if $G$ and $H$ are
non-isomorphic, there is always a differentiating property that Arthur can
verify in time polynomial in $n$.

But as noticed by Goldreich, Micali, and Wigderson \cite{gmw}, there is
something Merlin can do instead: he can let Arthur \textit{challenge} him.
\ Merlin can say:

\begin{quotation}
\noindent Arthur, send me a new graph $K$, which you obtained \textit{either}
by randomly permuting the vertices of $G$, \textit{or} randomly permuting the
vertices of $H$. \ Then I guarantee that I will tell you, without fail,
whether $K\cong G$\ or $K\cong H$.
\end{quotation}

It is clear that, if $G$ and $H$ are really non-isomorphic, then Merlin can
always answer such challenges correctly, by the assumption that he (Merlin)
has unlimited computational power. \ But it is equally clear that, if $G$ and
$H$ are isomorphic, then Merlin must answer some challenges incorrectly,
regardless of his computational power---since a random permutation of $G$ is
statistically indistinguishable from a random permutation of $H$.

This protocol has at least four features that merit reflection by anyone
interested in the nature of mathematical proof.

First, the protocol is \textit{probabilistic}. \ Merlin cannot convince Arthur
with certainty that $G$ and $H$ are non-isomorphic, since even if they were
isomorphic, there's a $1/2$\ probability that Merlin would get lucky and
answer a given challenge correctly (and hence, a $1/2^{k}$\ probability that
he would answer $k$ challenges correctly). \ All Merlin can do is offer to
repeat the protocol (say) $100$ or $1000$ times, and thereby make it less
likely that his proof is unsound than that an asteroid will strike Camelot,
killing both him and Arthur.

Second, the protocol is \textit{interactive}. \ Unlike with proof notions (1)
and (2), Arthur is no longer a passive recipient of knowledge, but an active
player who challenges the prover. \ We know from experience that the ability
to \textit{interrogate} a seminar speaker---to ask questions that the speaker
could not have anticipated, evaluate the responses, and then possibly ask
followup questions---often speeds up the process of figuring out whether the
speaker knows what he or she is talking about. \ Complexity theory affirms our
intuition here, through its discovery of interactive proofs for statements
(such as \textquotedblleft$G$ and $H$ are not isomorphic\textquotedblright)
whose shortest known conventional proofs are exponentially longer.

The third interesting feature of the graph non-isomorphism protocol---a
feature seldom mentioned---is that its soundness implicitly relies on a
\textit{physical} assumption. \ Namely, if Merlin had the power (whether
through magic or through ordinary espionage) to \textquotedblleft peer into
Arthur's study\textquotedblright\ and \textit{directly observe} whether Arthur
started with $G$ or $H$, then clearly he could answer every challenge
correctly even if $G\cong H$. \ It follows that the persuasiveness of Merlin's
\textquotedblleft proof\textquotedblright\ can only be as strong as Arthur's
extramathematical belief that Merlin does \textit{not} have such powers. \ By
now, there are many other examples in complexity theory of \textquotedblleft
proofs\textquotedblright\ whose validity rests on assumed limitations of the provers.

As Shieber \cite{shieber} points out, all three of the above properties of
interactive protocols \textit{also} hold for the Turing Test discussed in
Section \ref{AI}! \ The Turing Test is interactive by definition, it is
probabilistic because even a program that printed random gibberish would have
\textit{some} nonzero probability of passing the test by chance, and it
depends on the physical assumption that the AI program doesn't
\textquotedblleft cheat\textquotedblright\ by (for example) secretly
consulting a human. \ For these reasons, Shieber argues that we can see the
Turing Test \textit{itself} as an early interactive protocol---one that
convinces the verifier not of a mathematical theorem, but of the prover's
capacity for intelligent verbal behavior.\footnote{Incidentally, this provides
a good example of how notions from computational complexity theory can
influence philosophy even just at the level of metaphor, forgetting about the
actual results. \ In this essay, I didn't try to collect such
\textquotedblleft metaphorical\textquotedblright\ applications of complexity
theory, simply because there were too many of them!}

However, perhaps the most striking feature of the graph non-isomorphism
protocol is that it is \textit{zero-knowledge}:\ a technical term formalizing
our intuition that \textquotedblleft Arthur learns nothing from the
protocol,\ beyond the truth of the statement being proved.\textquotedblright%
\footnote{Technically, the protocol is \textquotedblleft%
\textit{honest-verifier} zero-knowledge,\textquotedblright\ meaning that
Arthur learns nothing from his conversation his Merlin besides the truth of
the statement being proved, \textit{assuming} Arthur follows the protocol
correctly. \ If Arthur cheats---for example, by sending a graph $K$ for which
he \textit{doesn't} already know an isomorphism either to $G$ or to $H$---then
Merlin's response could indeed tell Arthur something new. \ However,
Goldreich, Micali, and Wigderson \cite{gmw}\ also gave a more sophisticated
proof protocol for graph non-isomorphism, which remains zero-knowledge even in
the case where Arthur cheats.}\ \ For all Merlin ever tells Arthur is which
graph he (Arthur) started with, $G$ or $H$. \ But Arthur \textit{already knew}
which graph he started with! \ This means that, not only does Arthur gain no
\textquotedblleft understanding\textquotedblright\ of what makes $G$ and $H$
non-isomorphic, he does not even gain the ability to prove to a third party
what Merlin proved to him. \ This is another aspect of computational proofs
that has no analogue with proof notions (1) or (2).

One might complain that, as interesting as the zero-knowledge property is, so
far we've only shown it's achievable for an extremely specialized problem.
\ And indeed, just like with factoring integers, today there is strong
evidence that the graph isomorphism problem is \textit{not} $\mathsf{NP}%
$-complete \cite{bhz}.\footnote{Indeed, there is not even a consensus belief
that graph isomorphism is outside $\mathsf{P}$! \ The main reason is that, in
contrast to factoring integers, graph isomorphism turns out to be extremely
easy \textit{in practice}. \ Indeed, finding non-isomorphic graphs that
\textit{can't} be distinguished by simple invariants is itself a hard problem!
\ And in the past, several problems (such as linear programming and primality
testing) that were long known to be \textquotedblleft efficiently solvable for
practical purposes\textquotedblright\ were eventually shown to be
in\ $\mathsf{P}$\ in the strict mathematical sense as well.}$^{,}%
$\footnote{There is also strong evidence that there are short
\textit{conventional} proofs for graph non-isomorphism---in other words, that
not just graph isomorphism but also graph non-isomorphism will ultimately turn
out to be in $\mathsf{NP}$ \cite{kvm}.} \ However, in the same paper that gave
the graph non-isomorphism protocol, Goldreich, Micali, and Wigderson
\cite{gmw} also gave a celebrated zero-knowledge protocol (now called the
\textit{GMW protocol}) for the $\mathsf{NP}$-complete\ problems. \ By the
definition of $\mathsf{NP}$-complete\ (see Section \ref{ENTSCH}), the GMW
protocol meant that \textit{every mathematical statement that has a
conventional proof (say, in Zermelo-Fraenkel set theory) also has a
zero-knowledge proof of comparable size!} \ As an example application, suppose
you've just proved the Riemann Hypothesis. \ You want to convince the experts
of your triumph, but are paranoid about them stealing credit for it. \ In that
case, \textquotedblleft all\textquotedblright\ you need to do is

\begin{enumerate}
\item[(1)] rewrite your proof in a formal language,

\item[(2)] encode the result as the solution to an $\mathsf{NP}$-complete
problem, and then

\item[(3)] like a $16^{th}$-century court mathematician challenging his
competitors to a duel,\ invite the experts to run the GMW protocol with you
over the Internet!
\end{enumerate}

Provided you answer all their challenges correctly, the experts can become
\textit{statistically certain} that you possess a proof of the Riemann
Hypothesis, without learning anything \textit{about} that proof besides an
upper bound on its length.

Better yet, unlike the graph non-isomorphism protocol, the GMW protocol\ does
not assume a super-powerful wizard---only an ordinary polynomial-time being
who happens to know a proof of the relevant theorem. \ As a result, today the
GMW protocol is much more than a theoretical curiosity: it and its variants
have found major applications in Internet cryptography, where clients and
servers often need to prove to each other that they are following a protocol
correctly without revealing secret information as they do so.

However, there is one important caveat: unlike the graph-nonisomorphism
protocol, the GMW protocol relies essentially on a \textit{cryptographic
hypothesis}. \ For here is how the GMW protocol works: you (the prover) first
publish thousands of encrypted messages, each one \textquotedblleft
committing\textquotedblright\ you to a randomly-garbled piece of your claimed
proof. \ You then offer to decrypt a tiny fraction of those messages, as a way
for skeptical observers to \textquotedblleft spot-check\textquotedblright%
\ your proof, while learning nothing about its structure besides the useless
fact that, say, the $1729^{th}$\ step is valid (but how could it \textit{not}
be valid?). \ If the skeptics want to increase their confidence that your
proof is sound, then you simply run the protocol over and over with them,
using a fresh batch of encrypted messages each time. \ If the skeptics could
decrypt all the messages in a single batch, \textit{then} they could piece
together your proof---but to do that, they would need to break the underlying
cryptographic code.

\subsection{Other New Notions\label{OTHERNEW}}

Let me mention four other notions of \textquotedblleft proof\textquotedblright%
\ that complexity theorists have explored in depth over the last twenty years,
and that might merit philosophical attention.

\begin{itemize}
\item \textit{Multi-prover interactive proofs} \cite{bgkw,bfl}, in which
Arthur exchanges messages with \textit{two} (or more) computationally-powerful
but untrustworthy wizards. \ Here, Arthur might become convinced of some
mathematical statement, but only under the assumption that the wizards could
not communicate with \textit{each other} during the protocol. \ (The usual
analogy is to a police detective who puts two suspects in separate cells, to
prevent them from coordinating their answers.) \ Interestingly, in some
multi-prover protocols, even non-communicating wizards could successfully
coordinate their responses to Arthur's challenges (and thereby convince Arthur
of a falsehood) through the use of \textit{quantum entanglement}
\cite{chtw}.\ \ However, other protocols are conjectured to remain sound even
against entangled wizards \cite{kkmtv}.

\item \textit{Probabilistically checkable proofs} \cite{fglss,arorasafra},
which are mathematical proofs encoded in a special error-correcting format, so
that one can become confident of their validity by checking only $10$\ or
$20$\ bits chosen randomly in a correlated way. \ The \textit{PCP
(Probabilistically Checkable Proofs) Theorem} \cite{almss,dinur}, one of the
crowning achievements of complexity theory, says that \textit{any}
mathematical theorem, in any standard formal system such as Zermelo-Fraenkel
set theory, can be converted in polynomial time into a
probabilistically-checkable format.

\item \textit{Quantum proofs} \cite{watrous,ak}, which are proofs that depend
for their validity on the output of a quantum computation---possibly, even a
quantum computation that requires a special entangled \textquotedblleft proof
state\textquotedblright\ fed to it as input. \ Because $n$ quantum bits might
require $\sim2^{n}$\ classical bits to simulate, quantum proofs have the
property that it might never be possible to list all the \textquotedblleft
steps\textquotedblright\ that went into the proof, within the constraints of
the visible universe. \ For this reason, one's belief in the mathematical
statement being proved might depend on one's belief in the correctness of
quantum mechanics as a physical theory.

\item \textit{Computationally-sound proofs and arguments}\ \cite{bcc,micali},
which rely for their validity on the assumption that the prover was limited to
polynomial-time computations---as well as the mathematical conjecture that
crafting a convincing argument for a falsehood would have taken the prover
more than polynomial time.
\end{itemize}

What implications do these new types of proof have for the foundations of
mathematics? \ Do they merely make more dramatic what \textquotedblleft should
have been obvious all along\textquotedblright: that, as David Deutsch argues
in \textit{The Beginning of Infinity} \cite{deutsch:infinity}, proofs are
physical processes taking place in brains or computers, which therefore have
no validity independent of our beliefs about physics? \ Are the issues raised
essentially the same as those raised by \textquotedblleft
conventional\textquotedblright\ proofs that require extensive computations,
like Appel and Haken's proof of the Four-Color Theorem \cite{appelhaken}? \ Or
does appealing, in the course of a \textquotedblleft mathematical
proof,\textquotedblright\ to (say) the validity of quantum mechanics, the
randomness of apparently-random numbers, or the lack of certain superpowers on
the part of the prover represent something qualitatively new? \ Philosophical
analysis is sought.

\section{Complexity, Space, and Time\label{TIME}}

What can computational complexity tell us about the nature of space and time?
\ A first answer might be \textquotedblleft not much\textquotedblright: after
all, the definitions of standard complexity classes such as $\mathsf{P}$\ can
be shown to be insensitive to such details as the number of spatial
dimensions, and even whether the speed of light is finite or
infinite.\footnote{More precisely, Turing machines with one-dimensional tapes
are polynomially equivalent to Turing machines with $k$-dimensional tapes for
any $k$, and are also polynomially equivalent to \textit{random-access
machines} (which can \textquotedblleft jump\textquotedblright\ to any memory
location in unit time, with no locality constraint).
\par
On the other hand, if we care about polynomial differences in speed, and
\textit{especially} if we want to study parallel computing models, details
about the spatial layout of the computing and memory elements (as well as the
speed of communication among the elements) can become vitally important.} \ On
the other hand, I think complexity theory does offer insight about the
\textit{differences} between space and time.

The class of problems solvable using a polynomial amount of memory (but
possibly an exponential amount of time\footnote{Why \textquotedblleft
only\textquotedblright\ an exponential amount? \ Because a Turing machine with
$B$\ bits of memory can run for no more than $2^{B}$\ time steps. \ After
that, the machine must either halt or else return to a configuration
previously visited (thereby entering an infinite loop).}) is called
$\mathsf{PSPACE}$, for Polynomial Space. \ Examples of $\mathsf{PSPACE}%
$\ problems include simulating dynamical systems, deciding whether a regular
grammar generates all possible strings, and executing an optimal strategy in
two-player games such as Reversi, Connect Four, and
Hex.\footnote{\label{chessnote}Note that, in order to speak about the
computational complexity of such games, we first need to generalize them to an
$n\times n$\ board! \ But if we do so, then for many natural games, the
problem of determining which player has the win from a given position is not
only in $\mathsf{PSPACE}$, but $\mathsf{PSPACE}$\textit{-complete} (i.e., it
captures the entire difficulty of the class $\mathsf{PSPACE}$). \ For example,
Reisch \cite{reisch} showed that this is true for Hex.
\par
What about a suitable\ generalization of \textit{chess} to an $n\times
n$\ board? \ That's also in $\mathsf{PSPACE}$---but as far as anyone knows,
only if we impose a polynomial upper bound on the number of moves in a chess
game. \ Without such a restriction, Fraenkel and Lichtenstein \cite{fraenkel}
showed that chess is $\mathsf{EXP}$-complete; with such a restriction, Storer
\cite{storer}\ showed that chess is $\mathsf{PSPACE}$-complete.} \ It is not
hard to show that $\mathsf{PSPACE}$\ is at least as powerful as $\mathsf{NP}$:%
\[
\mathsf{P}\subseteq\mathsf{NP}\subseteq\mathsf{PSPACE}\subseteq\mathsf{EXP}.
\]
Here $\mathsf{EXP}$\ represents the class of problems solvable using an
exponential amount of time, and also possibly an exponential amount of
memory.\footnote{In this context, we call a function $f\left(  n\right)
$\ \textquotedblleft exponential\textquotedblright\ if it can be upper-bounded
by $2^{p\left(  n\right)  }$, for some polynomial $p$. \ Also, note that
\textit{more} than exponential memory would be useless here, since a Turing
machine that runs for $T$ time steps can visit at most $T$\ memory cells.}
\ Every one of the above containments is believed to be strict, although the
only one currently \textit{proved} to be strict is $\mathsf{P}\neq
\mathsf{EXP}$, by an important 1965 result of Hartmanis and Stearns
\cite{hs}\ called the Time Hierarchy Theorem\footnote{More generally, the Time
Hierarchy Theorem shows that, if $f$\ and $g$\ are any two \textquotedblleft
sufficiently well-behaved\textquotedblright\ functions that satisfy $f\left(
n\right)  \ll g\left(  n\right)  $ (for example: $f\left(  n\right)  =n^{2}%
$\ and $g\left(  n\right)  =n^{3}$), then \textit{there are computational
problems solvable in }$g\left(  n\right)  $\textit{\ time but not in
}$f\left(  n\right)  $\textit{\ time}. \ The proof of this theorem uses
diagonalization, and can be thought of as a scaled-down version of Turing's
proof of the unsolvability of the halting problem. \ That is, we argue that,
\textit{if} it were always possible to simulate a $g\left(  n\right)  $-time
Turing machine by an $f\left(  n\right)  $-time Turing machine, then we could
construct a $g\left(  n\right)  $-time machine that \textquotedblleft
predicted its own output in advance\textquotedblright\ and then output
something else---thereby causing a contradiction.
\par
Using similar arguments, we can show (for example) that there exist
computational problems solvable using $n^{3}$\ bits of memory but not using
$n^{2}$\ bits, and so on in most cases where we want to compare \textit{more
versus less of the same computational resource.} \ In complexity theory, the
hard part is comparing two \textit{different} resources: for example,
determinism versus nondeterminism (the $\mathsf{P}\overset{?}{=}\mathsf{NP}%
$\ problem), time versus space ($\mathsf{P\overset{?}{=}PSPACE}$), or
classical versus quantum computation ($\mathsf{BPP\overset{?}{=}BQP}$). \ For
in those cases, diagonalization by itself no longer works.}$^{,}$\footnote{The
fact that $\mathsf{P}\neq\mathsf{EXP}$\ has an amusing implication, often
attributed to Hartmanis: namely, \textit{at least one} of the three
inequalities
\par
\begin{enumerate}
\item[(i)] $\mathsf{P}\neq\mathsf{NP}$
\par
\item[(ii)] $\mathsf{NP}\neq\mathsf{PSPACE}$
\par
\item[(iii)] $\mathsf{PSPACE}\neq\mathsf{EXP}$
\end{enumerate}
\par
\noindent must be true, even though proving any one of them to be true
\textit{individually} would represent a titanic advance in mathematics!
\par
The above\ observation is sometimes offered as circumstantial evidence for
$\mathsf{P}\neq\mathsf{NP}$. Of all our hundreds of unproved beliefs about
inequalities between pairs of complexity classes, a large fraction of them
\textit{must} be correct, simply to avoid contradicting the hierarchy
theorems. \ So then why not $\mathsf{P}\neq\mathsf{NP}$\ in particular (given
that our intuition there is stronger than our intuitions for most of the other
inequalities)?}.

Notice, in particular, that $\mathsf{P}\neq\mathsf{NP}$ implies $\mathsf{P\neq
PSPACE}$.\ \ So while $\mathsf{P\neq PSPACE}$\ is not yet proved, it is an
extremely secure conjecture by the standards of complexity theory. \ In slogan
form, complexity theorists believe that \textit{space is more powerful than
time}.

Now, some people have asked how such a claim could possibly be consistent with
modern physics. \ For didn't Einstein teach us that space and time are merely
two aspects of the same structure? \ One immediate answer is that, even
\textit{within} relativity theory, space and time are not interchangeable:
space has a positive signature\ whereas time has a negative signature. \ In
complexity theory, the difference between space and time manifests itself in
the straightforward fact that you can \textit{reuse} the same memory cells
over and over, but you can't reuse the same moments of time.\footnote{See my
blog post www.scottaaronson.com/blog/?p=368\ for more on this theme.}

Yet, as trivial as that observation sounds, it leads to an interesting
thought. \ Suppose that the laws of physics let us travel \textit{backwards
}in time. \ In such a case, it's natural to imagine that time would become a
\textquotedblleft reusable resource\textquotedblright\ just like space
is---and that, as a result, arbitrary $\mathsf{PSPACE}$\ computations would
fall within our grasp. \ But is that just an idle speculation, or can we
rigorously justify it?

\subsection{Closed Timelike Curves\label{CTC}}

Philosophers, like science-fiction fans, have long been interested in the
possibility of closed timelike curves (CTCs), which arise in certain solutions
to Einstein's field equations of general relativity.\footnote{Though it is not
known whether those solutions are \textquotedblleft physical\textquotedblright%
: for example, whether or not they can survive in a quantum theory of gravity
(see \cite{mty} for example).} \ On a traditional understanding, the central
philosophical problem raised by CTCs is the \textit{grandfather paradox}.
\ This is the situation where you go back in time to kill your own
grandfather, therefore you are never born, therefore your grandfather is
\textit{not} killed, therefore you \textit{are} born, and so on. \ Does this
contradiction immediately imply that\ CTCs are impossible?

No, it doesn't: we can only conclude that, \textit{if} CTCs exist, then the
laws of physics must somehow prevent grandfather paradoxes from arising. \ How
could they do so? \ One classic illustration is that \textquotedblleft when
you go back in time to try and kill your grandfather, the gun
jams\textquotedblright---or some other \textquotedblleft
unlikely\textquotedblright\ event inevitably occurs to keep the state of the
universe consistent. \ But why should we imagine that such a convenient
\textquotedblleft out\textquotedblright\ will always be available, in every
physical experiment involving CTCs? \ Normally, we like to imagine that we
have the freedom to design an experiment however we wish, without Nature
imposing conditions on the experiment (for example: \textquotedblleft every
gun must jam sometimes\textquotedblright) whose reasons can only be understood
in terms of distant or hypothetical events.

In his 1991 paper \textquotedblleft Quantum mechanics near closed timelike
lines,\textquotedblright\ Deutsch \cite{deutsch:ctc} gave an elegant proposal
for eliminating grandfather paradoxes. \ In particular he showed that, as long
as we assume the laws of physics are quantum-mechanical (or even just
classically probabilistic), every experiment involving a CTC admits at least
one \textit{fixed point}: that is, a way to satisfy the conditions of the
experiment that ensures consistent evolution. \ Formally, if $S$ is the
mapping from quantum states to themselves induced by \textquotedblleft going
around the CTC once,\textquotedblright\ then a fixed point is any quantum
mixed state\footnote{In quantum mechanics, a \textit{mixed state} can be
thought of as a classical probability distribution over quantum states.
\ However, an important twist is that the same mixed state can be represented
by \textit{different} probability distributions: for example, an equal mixture
of the states $\left\vert 0\right\rangle $\ and $\left\vert 1\right\rangle
$\ is physically indistinguishable from an equal mixture of $\frac{\left\vert
0\right\rangle +\left\vert 1\right\rangle }{\sqrt{2}}$\ and $\frac{\left\vert
0\right\rangle -\left\vert 1\right\rangle }{\sqrt{2}}$. \ This is why mixed
states are represented mathematically using Heisenberg's density matrix
formalism.} $\rho$\ such that $S\left(  \rho\right)  =\rho$. \ The existence
of such a $\rho$\ follows from simple linear-algebraic arguments. \ As one
illustration, the \textquotedblleft resolution of the grandfather
paradox\textquotedblright\ is now that you are born with probability $1/2$,
and \textit{if} you are born, you go back in time to kill your
grandfather---from which it follows that you are born with probability $1/2$,
and so on. \ Merely by treating states as probabilistic (as, in some sense,
they \textit{have} to be in quantum mechanics\footnote{In more detail,
Deutsch's proposal works if the state space consists of classical probability
distributions $\mathcal{D}$\ or quantum mixed states $\rho$, but \textit{not}
if it consists of pure states $\left\vert \psi\right\rangle $. \ Thus,
\textit{if} one believed that only pure states were fundamental in physics,
and that probability distributions and mixed states always reflected
subjective ignorance, one might reject Deutsch's proposal on that ground.}),
we have made the evolution of the universe consistent.

But Deutsch's account of CTCs faces at least three serious difficulties. \ The
first difficulty is that the fixed points might not be \textit{unique}: there
could be many mixed states $\rho$\ such that $S\left(  \rho\right)  =\rho$,
and then the question arises of how Nature chooses one of them. \ To
illustrate, consider the \textit{grandfather anti-paradox}: a bit
$b\in\left\{  0,1\right\}  $\ that travels around a CTC without changing. \ We
can consistently assume $b=0$, or $b=1$, or any probabilistic mixture of the
two---and unlike the usual situation in physics, here there is no possible
boundary condition that could resolve the ambiguity.

The second difficulty, pointed out Bennett et al.\ \cite{blss},\ is that
Deutsch's proposal violates the statistical interpretation of quantum mixed
states. So for example, if half of an entangled pair%
\[
\frac{\left\vert 0\right\rangle _{A}\left\vert 0\right\rangle _{B}+\left\vert
1\right\rangle _{A}\left\vert 1\right\rangle _{B}}{\sqrt{2}}%
\]
is placed inside the CTC, while the other half remains outside the CTC, then
the process of finding a fixed point will \textquotedblleft
break\textquotedblright\ the entanglement between the two halves. \ As a
\textquotedblleft remedy\textquotedblright\ for this problem, Bennett et
al.\ suggest requiring the CTC fixed point $\rho$\ to be independent of the
entire rest of the universe. \ To my mind, this remedy is so drastic that it
basically amounts to defining CTCs out of existence!

Motivated by these difficulties, Lloyd et al.\ \cite{lmggs}\ recently proposed
a completely different account of CTCs, based on \textit{postselected
teleportation}. \ Lloyd et al.'s account avoids both of the problems
above---though perhaps not surprisingly, introduces other problems of its
own.\footnote{In particular, in Lloyd et al.'s proposal, the only way to deal
with the grandfather paradox is by some variant of \textquotedblleft the gun
jams\textquotedblright: there \textit{are} evolutions with no consistent
solution, and it needs to be postulated that the laws of physics are such that
they never occur.}\ \ My own view, for whatever it is worth, is that Lloyd et
al.\ are talking less about \textquotedblleft true\textquotedblright\ CTCs as
I would understand the concept, as about postselected quantum-mechanical
experiments that \textit{simulate} CTCs in certain interesting respects. \ If
there are any controversies in physics that call out for expert philosophical
attention, surely this is one of them.

\subsection{The Evolutionary Principle\label{EP}}

Yet so far, we have not even mentioned what I see as the \textit{main}
difficulty with Deutsch's account of CTCs. \ This is that \textit{finding} a
fixed point might require Nature to solve an astronomically-hard computational
problem! \ To illustrate, consider a science-fiction scenario wherein you go
back in time and dictate Shakespeare's plays to him. \ Shakespeare thanks you
for saving him the effort, publishes verbatim the plays that you dictated, and
centuries later the plays come down to you, whereupon you go back in time and
dictate them to Shakespeare, etc.

Notice that, in contrast to the grandfather paradox, here there is no logical
contradiction: the story as we told it is entirely consistent. \ But most
people find the story \textquotedblleft paradoxical\textquotedblright\ anyway.
\ After all, somehow \textit{Hamlet} gets written, without anyone ever doing
the work of writing it! \ As Deutsch \cite{deutsch:ctc} perceptively observed,
if there is a \textquotedblleft paradox\textquotedblright\ here, then it is
not one of logic but of \textit{computational complexity}. \ Specifically, the
story violates a commonsense principle that we can loosely articulate as
follows:%
\[
\text{\textbf{Knowledge requires a causal process to bring it into
existence.}}%
\]
Like many other important principles, this one might not be recognized as a
\textquotedblleft principle\textquotedblright\ at all before we contemplate
situations that violate it! \ Deutsch \cite{deutsch:ctc}\ calls this principle
the \textit{Evolutionary Principle} (EP). \ Note that some version of the EP
was invoked both by William Paley's blind-watchmaker argument, and
(ironically) by the arguments of Richard Dawkins \cite{dawkins}\ and other
atheists against the existence of an intelligent designer.

In my survey article \textquotedblleft$\mathsf{NP}$-Complete Problems and
Physical Reality\textquotedblright\ \cite{aar:np}, I proposed and defended a
complexity-theoretic analogue of the EP, which I called the $\mathsf{NP}%
$\ \textit{Hardness Assumption}:%
\[
\text{\textbf{There is no physical means to solve }}\mathsf{NP}%
\text{\textbf{-complete problems in polynomial time.}}%
\]
The above statement implies $\mathsf{P}\neq\mathsf{NP}$, but is stronger in
that it encompasses probabilistic computing, quantum computing, and
\textit{any other} \textit{computational model} compatible with the laws of
physics. \ See \cite{aar:np}\ for a survey of recent results bearing on the
$\mathsf{NP}$ Hardness Assumption,\ analyses of claimed counterexamples to the
assumption, and possible implications of the assumption for physics.

\subsection{Closed Timelike Curve Computation\label{CTCCOMP}}

But can we show more rigorously that closed timelike curves would
\textit{violate} the $\mathsf{NP}$ Hardness Assumption? \ Indeed, let us now
show that, in a universe where arbitrary computations could be performed
inside a CTC, and where Nature had to find a fixed point for the CTC, we could
solve $\mathsf{NP}$-complete problems using only polynomial resources.

We can model any $\mathsf{NP}$-complete problem instance by a function
$f:\left\{  0,\ldots,2^{n}-1\right\}  \rightarrow\left\{  0,1\right\}  $,
which maps each possible solution $x$\ to the bit $1$\ if $x$ is valid, or to
$0$\ if $x$ is invalid. \ (Here, for convenience, we identify each $n$-bit
solution string $x$\ with the\ nonnegative integer that $x$\ encodes in
binary.) \ Our task, then, is to find an $x\in\left\{  0,\ldots,2^{n}%
-1\right\}  $\ such that $f\left(  x\right)  =1$. \ We can solve this problem
with just a \textit{single} evaluation to $f$, provided we can run the
following computer program $C$\ inside a closed timelike curve
\cite{brun,aar:np,awat}:

\begin{quotation}
\texttt{Given input }$x\in\left\{  0,\ldots,2^{n}-1\right\}  $\texttt{:}

\texttt{If }$f\left(  x\right)  =1$\texttt{, then output }$x$

\texttt{Otherwise, output }$\left(  x+1\right)  \operatorname{mod}2^{n}$
\end{quotation}

Assuming there exists at least one $x$ such that $f\left(  x\right)  =1$, the
only \textit{fixed points} of $C$---that is, the only ways for $C$'s output to
equal its input---are for $C$ to input, and output, such a valid solution $x$,
which therefore appears in $C$'s output register \textquotedblleft as if by
magic.\textquotedblright\ \ (If there are no valid solutions,\ then $C$'s
fixed points will simply be uniform superpositions or probability
distributions over \textit{all} $x\in\left\{  0,\ldots,2^{n}-1\right\}  $.)

Extending the above idea, John Watrous and I \cite{awat}\ (following a
suggestion by Fortnow) recently showed that a CTC computer in Deutsch's model
could solve all problems in $\mathsf{PSPACE}$. \ (Recall that $\mathsf{PSPACE}%
$\ is believed to be even larger than $\mathsf{NP}$.) \ More surprisingly, we
also showed that $\mathsf{PSPACE}$\ constitutes the \textit{limit} on what can
be done with a CTC computer; and that this is true whether the CTC computer is
classical or quantum. \ One consequence of our results is that the
\textquotedblleft na\"{\i}ve intuition\textquotedblright\ about CTC
computers---that their effect would be to \textquotedblleft make space and
time equivalent as computational resources\textquotedblright---is ultimately
correct, although not for the na\"{\i}ve\ reasons.\footnote{Specifically, it
is \textit{not} true that in a CTC universe, a Turing machine tape head could
just travel back and forth in time the same way it travels back and forth in
space. \ If one thinks this way, then one really has in mind a second,
\textquotedblleft meta-time,\textquotedblright\ while the \textquotedblleft
original\textquotedblright\ time has become merely one more dimension of
space. \ To put the point differently: even though a CTC would make time
\textit{cyclic}, time would still retain its \textit{directionality}. \ This
is the reason why, if we want to show that CTC computers have the power of
$\mathsf{PSPACE}$, we need a nontrivial argument involving causal
consistency.} \ A second, amusing consequence is that, once closed timelike
curves are available, switching from classical to quantum computers provides
no \textit{additional} benefit!

It is important to realize that our algorithms for solving hard problems with
CTCs do \textit{not} just boil down to \textquotedblleft using huge amounts of
time to find the answer, then sending the answer back in time to before the
computer started.\textquotedblright\ \ For even in the exotic scenario of a
time travel computer, we still require that all resources used \textit{inside}
the CTC (time, memory, etc.) be polynomially-bounded. \ Thus, the ability to
solve hard problems comes solely from \textit{causal consistency}: the
requirement that Nature must find some evolution for the CTC computer that
avoids grandfather paradoxes.

In Lloyd et al.'s alternative account of CTCs based on postselection
\cite{lmggs}, hard problems can \textit{also} be solved, though for different
reasons. \ In particular, building on an earlier result of mine \cite{aar:pp},
Lloyd et al.\ show that the power of their model corresponds to a complexity
class called $\mathsf{PP}$ (Probabilistic Polynomial-Time), which is believed
to be strictly smaller than $\mathsf{PSPACE}$ but strictly larger than
$\mathsf{NP}$. \ Thus, one might say that Lloyd et al.'s model
\textquotedblleft improves\textquotedblright\ the computational situation, but
not by much!

So one might wonder: is there any way that the laws of physics could allow
CTCs, \textit{without} opening the door to implausible computational powers?
\ There remains at least one interesting possibility, which was communicated
to me by the philosopher Tim Maudlin.\footnote{This possibility is also
discussed at length in Deutsch's paper \cite{deutsch:ctc}.} \ Maybe the laws
of physics have the property that, no matter what computations are performed
inside a CTC, Nature always has an \textquotedblleft out\textquotedblright%
\ that avoids the grandfather paradox, but \textit{also} avoids solving hard
computational problems---analogous to \textquotedblleft the gun
jamming\textquotedblright\ in the original grandfather paradox. \ Such an
out\ might involve (for example) an asteroid hitting the CTC computer, or the
computer failing for other mysterious reasons. \ Of course, \textit{any}
computer in the physical world has some nonzero probability of failure, but
ordinarily\ we imagine that the failure probability can be made negligibly
small. \ However, in situations where Nature is being \textquotedblleft
forced\textquotedblright\ to find a fixed point, maybe \textquotedblleft
mysterious computer failures\textquotedblright\ would become the norm rather
than the exception.

To summarize, I think that computational complexity theory \textit{changes}
the philosophical issues raised by time travel into the past. \ While
discussion traditionally focused on the grandfather paradox, we have seen that
there is no shortage of ways for Nature to avoid logical inconsistencies, even
in a universe with CTCs. \ The \textquotedblleft real\textquotedblright%
\ problem, then, is how to escape the \textit{other} paradoxes that arise in
the course of taming the grandfather paradox! \ Probably foremost among those
is the \textquotedblleft computational complexity paradox,\textquotedblright%
\ of $\mathsf{NP}$-complete\ and even harder problems getting solved as if by magic.

\section{Economics\label{ECON}}

In classical economics, agents are modeled as rational, Bayesian agents who
take whatever actions will maximize their expected utility $\operatorname*{E}%
_{\omega\in\Omega}\left[  U\left(  \omega\right)  \right]  $, given their
subjective probabilities $\left\{  p_{\omega}\right\}  _{\omega\in\Omega}%
$\ over all possible states $\omega$\ of the world.\footnote{Here we assume
for simplicity that the set $\Omega$\ of possible states is countable;
otherwise we could of course use a continuous probability measure.} \ This, of
course, is a caricature that seems almost designed to be attacked, and it
\textit{has} been attacked from almost every angle. \ For example, humans are
not even close to rational Bayesian agents, but suffer from well-known
cognitive biases, as explored by Kahneman and Tversky \cite{kahnemantversky}%
\ among others. \ Furthermore, the classical view seems to leave no room for
critiquing people's beliefs (i.e., their prior probabilities) or their utility
functions as irrational---yet it is easy to cook up prior probabilities or
utility functions that would lead to behavior that almost anyone would
consider insane. \ A third problem is that, in games with several cooperating
or competing agents who act simultaneously, classical economics guarantees the
existence of at least one \textit{Nash equilibrium} among the agents'
strategies. \ But the usual situation is that there are multiple equilibria,
and then there is no general principle to predict which equilibrium will
prevail, even though the choice might mean the difference between war and peace.

Computational complexity theory can contribute to debates about the
foundations of economics by showing that, even in the idealized situation of
rational agents who all have perfect information about the state of the world,
it will often be \textit{computationally intractable} for those agents to act
in accordance with\ classical economics. \ Of course, some version of this
observation has been recognized in economics for a long time. \ There is a
large literature on \textit{bounded rationality} (going back to the work of
Herbert Simon \cite{herbsimon}), which studies the behavior of economic agents
whose decision-making abilities are limited in one way or another.

\subsection{Bounded Rationality and the Iterated Prisoners' Dilemma\label{IPD}%
}

As one example of an insight to emerge from this literature, consider the
Finite Iterated Prisoner's Dilemma. \ This is a game where two players meet
for some fixed number of rounds $N$, which is finite and common knowledge
between the players. \ In each round, both players can either
\textquotedblleft Defect\textquotedblright\ or \textquotedblleft
Cooperate\textquotedblright\ (not knowing the other player's choice), after
which they receive the following payoffs:%
\[%
\begin{tabular}
[c]{lll}
& Defect$_{2}$ & Cooperate$_{2}$\\
Defect$_{1}$ & $1,1$ & $4,0$\\
Cooperate$_{1}$ & $0,4$ & $3,3$%
\end{tabular}
\ \
\]
Both players remember the entire previous history of the interaction. \ It is
clear that the players will be jointly best off if they both cooperate, but
equally clear that if $N=1$, then cooperation is not an equilibrium. \ On the
other hand,\ \textit{if the number of rounds }$N$\textit{ were unknown or
infinite}, then the players could rationally decide to cooperate, similarly to
how humans decide to cooperate in real life. \ That is, Player 1 reasons that
if he defects, then Player 2 will retaliate by defecting in future rounds, and
vice versa. \ So over the long run, both players do best for themselves by cooperating.

The \textquotedblleft paradox\textquotedblright\ is now that, as soon as $N$
becomes known, the above reasoning collapses. \ For assuming the players are
rational, they both realize that whatever else, neither has anything to lose
by defecting \textit{in round} $N$---and therefore that is what they do. \ But
since both players \textit{know} that both will defect in round $N$, neither
one has anything to lose by defecting in round $N-1$ \textit{either}---and
they can continue inductively in this way back to the first round. \ We
therefore get the \textquotedblleft prediction\textquotedblright\ that both
players will defect in every round, even though that is neither in the
players' own interests, nor what actual humans do in experiments.

In 1985, Neyman \cite{neyman} proposed an ingenious resolution of this
paradox. \ Specifically, he showed that if the two players have
\textit{sufficiently small memories}---technically, if they are finite
automata with $k$\ states, for $2\leq k<N$---then cooperation becomes an
equilibrium once again! \ The basic intuition is that, if both players lack
enough memory to count up to $N$, and both of them know that, and both know
that they both know that, and so on, then the inductive argument in the last
paragraph fails, since it assumes intermediate strategies that neither player
can implement.

While complexity considerations vanquish \textit{some} of
the\ counterintuitive conclusions of classical economics, equally interesting
to me is that they do not vanquish others. \ As one example, I showed in
\cite{aar:agr}\ that Robert Aumann's celebrated \textit{agreement theorem}
\cite{aumann}---perfect Bayesian agents with common priors can never
\textquotedblleft agree to disagree\textquotedblright---persists even in the
presence of limited communication between the agents.

There are many other interesting results in the bounded rationality
literature, too many to do them justice here (but see Rubinstein
\cite{rubinstein:book}\ for a survey). \ On the other hand, \textquotedblleft
bounded rationality\textquotedblright\ is something of a catch-all phrase,
encompassing almost every imaginable deviation from rationality---including
human cognitive biases, limits on information-gathering and communication, and
the\ restriction of strategies to a specific form (for example, linear
threshold functions). \ Many of these deviations have little to do with
computational complexity \textit{per se}. \ So the question remains of whether
computational complexity \textit{specifically} can provide new insights about
economic behavior.

\subsection{The Complexity of Equilibria\label{EQUILIB}}

There are some very recent advances suggesting that the answer is yes.
\ Consider the problem of finding an equilibrium of a two-player game, given
the $n\times n$ payoff matrix as input. \ In the special case of
\textit{zero-sum games} (which von Neumann studied in 1928), it has long been
known how to solve this problem in an amount of time polynomial in $n$, for
example by reduction to linear programming. \ But in 2006, Daskalakis,
Goldberg, and Papadimitriou \cite{dgp}\ (with improvements by Chen and Deng
\cite{chendeng}) proved the spectacular result that, for a \textit{general}
(not necessarily zero-sum) two-player game, finding a Nash equilibrium is
\textquotedblleft$\mathsf{PPAD}$-complete.\textquotedblright\ \ Here
$\mathsf{PPAD}$\ (\textquotedblleft Polynomial Parity Argument,
Directed\textquotedblright) is, roughly speaking, the class of \textit{all}
search problems for which a solution is guaranteed to exist for the same
combinatorial reason that every game has at least one Nash equilibrium. \ Note
that finding a Nash equilibrium \textit{cannot} be $\mathsf{NP}$%
-complete,\ for the technical reason that $\mathsf{NP}$ is a class of
\textit{decision }problems, and the answer to the decision problem
\textquotedblleft does this game have a Nash equilibrium?\textquotedblright%
\ is always yes. \ But Daskalakis et al.'s result says (informally) that the
search problem of \textit{finding} a Nash problem is \textquotedblleft as
close to $\mathsf{NP}$-complete as it could possibly be,\textquotedblright%
\ subject to its decision version being trivial. \ Similar $\mathsf{PPAD}%
$-completeness\ results are now known for other fundamental economic problems,
such as finding market-clearing prices in Arrow-Debreu markets \cite{cddt}.

Of course, one can debate the economic relevance of these results: for
example, how often does the computational hardness that we now
know\footnote{Subject, as usual, to widely-believed complexity assumptions.}
to be inherent in economic equilibrium theorems actually rear its head in
practice? \ But one can similarly debate the economic relevance of the
equilibrium theorems themselves! \ In my opinion, if the theorem that Nash
equilibria \textit{exist} is considered relevant to debates about (say) free
markets versus government intervention, then the theorem that \textit{finding}
those equilibria is $\mathsf{PPAD}$-complete should be considered relevant also.

\section{Conclusions\label{CONC}}

The purpose of this essay was to illustrate how philosophy could be enriched
by taking computational complexity theory into account, much as it was
enriched almost a century ago by taking computability theory into account.
\ In particular, I argued that computational complexity provides new insights
into the explanatory content of Darwinism, the nature of mathematical
knowledge and proof, computationalism, syntax versus semantics, the problem of
logical omniscience, debates surrounding the Turing Test and Chinese Room, the
problem of induction, the foundations of quantum mechanics, closed timelike
curves, and economic rationality.

Indeed, one might say that the \textquotedblleft real\textquotedblright%
\ question is which philosophical problems \textit{don't} have important
computational complexity aspects! \ My own opinion is that there probably
\textit{are} such problems (even within analytic philosophy), and that one
good candidate is the problem of what we should take as \textquotedblleft
bedrock mathematical reality\textquotedblright: that is, the set of
mathematical statements that are objectively true or false, regardless of
whether they can be proved or disproved in a given formal system. \ To me, if
we are not willing to say that a given Turing machine $M$\ either accepts,
rejects, or runs forever (when started on a blank tape)---and that which one
it does is an objective fact, independent of our formal axiomatic theories,
the laws of physics, the biology of the human brain, cultural conventions,
etc.---then \textit{we have no basis to talk about any of those other things}
(axiomatic theories, the laws of physics, and so on). \ Furthermore, $M$'s
resource requirements are irrelevant here: even if $M$\ only halts after
$2^{2^{10000}}$\ steps, its output is as mathematically definite as if it had
halted after $10$ steps.\footnote{The situation is very different for
mathematical statements like the Continuum Hypothesis, which \textit{can't}
obviously be phrased as predictions about idealized computational processes
(since they're not expressible by first-order or even second-order
quantification over the integers). \ For those statements, it really
\textit{is} unclear to me what one means by their truth or falsehood apart
from their provability in some formal system.}

Can we say anything \textit{general} about when a computational complexity
perspective is helpful in philosophy, and when it isn't? \ Extrapolating from
the examples in this essay, I would say that computational complexity tends to
be helpful when we want to know whether a particular fact \textit{does any
explanatory work}:\ Sections \ref{EVOL}, \ref{INTEGERS}, \ref{AI},
\ref{WATERFALLS}, and \ref{PAC}\ all provided examples of this. \ Other
\textquotedblleft philosophical\ applications\textquotedblright\ of complexity
theory come from the Evolutionary Principle and the $\mathsf{NP}$ Hardness
Assumption discussed in Section \ref{EP}. \ \textit{If} we believe that
certain problems are computationally intractable, then we may be able to draw
interesting conclusions from that belief about economic rationality, quantum
mechanics, the possibility of closed timelike curves, and other issues. \ By
contrast, computational complexity tends to be \textit{un}helpful when we only
want to know whether a particular fact \textquotedblleft
determines\textquotedblright\ another fact, and don't care about the length of
the inferential chain.

\subsection{Criticisms of Complexity Theory\label{CRITICISMS}}

Despite its explanatory reach, complexity theory has been criticized on
various grounds. \ Here are four of the most common criticisms:

\begin{enumerate}
\item[(1)] Complexity theory only makes \textit{asymptotic} statements
(statements about how the resources needed to solve problem instances of size
$n$ scale as $n$ goes to infinity). \ But as a matter of logic, asymptotic
statements need not have \textit{any implications whatsoever} for the finite
values of $n$ (say, $10,000$) that humans care actually about, nor can any
finite amount of experimental data confirm or refute an asymptotic claim.

\item[(2)] Many of (what we would like to be) complexity theory's basic
principles, such as $\mathsf{P}\neq\mathsf{NP}$, are currently unproved
mathematical conjectures, and will probably remain that way for a long time.

\item[(3)] Complexity theory focuses on only a limited type of computer---the
serial, deterministic Turing machine---and fails to incorporate the
\textquotedblleft messier\textquotedblright\ computational phenomena found in nature.

\item[(4)] Complexity theory studies only the \textit{worst-case} behavior of
algorithms, and does not address whether that behavior is representative, or
whether it merely reflects a few \textquotedblleft
pathological\textquotedblright\ inputs. \ So for example, even if\ $\mathsf{P}%
\neq\mathsf{NP}$, there might still be excellent heuristics to solve
\textit{most} instances of $\mathsf{NP}$-complete problems that actually arise
in practice; complexity theory tells us nothing about such possibilities one
way or the other.
\end{enumerate}

For whatever it's worth, criticisms (3) and (4) have become much less accurate
since the 1980s. \ As discussed in this essay, complexity theory has by now
branched out far beyond deterministic Turing machines, to incorporate (for
example) quantum mechanics, parallel and distributed computing, and stochastic
processes such as Darwinian evolution. \ Meanwhile, although worst-case
complexity remains the best-understood kind, today there is a large body of
work---much of it driven by cryptography---that studies the
\textit{average-case} hardness of computational problems, for various
probability distributions over inputs. \ And just as almost all complexity
theorists believe that $\mathsf{P}\neq\mathsf{NP}$, so almost all subscribe to
the stronger belief that there exist \textit{hard-on-average} $\mathsf{NP}$
problems---indeed, that belief is one of the underpinnings of modern
cryptography. \ A few problems, such as calculating discrete logarithms, are
even known to be \textit{just as hard on random inputs as they are on the
hardest possible input} (though whether such \textquotedblleft
worst-case/average-case equivalence\textquotedblright\ holds for any
$\mathsf{NP}$-complete problem remains a major open question). \ For these
reasons, although speaking about average-case rather than worst-case
complexity would complicate some of the arguments in this essay, I don't think
it would change the conclusions much.\footnote{On the other hand, it
\textit{would} presuppose that we knew how to define reasonable probability
distributions over inputs. \ But as discussed in Section \ref{HUMANS},\ it
seems hard to explain what we mean by\ \textquotedblleft structured
instances,\textquotedblright\ or \textquotedblleft the types of instances that
normally arise in practice.\textquotedblright} \ See Bogdanov and Trevisan
\cite{bt} for an excellent recent survey of average-case complexity, and
Impagliazzo \cite{impagliazzo} for an evocative discussion of complexity
theory's \textquotedblleft possible worlds\textquotedblright\ (for example,
the \textquotedblleft world\textquotedblright\ where $\mathsf{NP}$-complete
problems\ turn out to be hard in the worst case but easy on average).

The broader point is that, even if we admit that criticisms (1)-(4) have
merit, that does not give us a license to dismiss complexity-theoretic
arguments whenever we dislike them! \ In science, we only ever deal with
imperfect, approximate theories---and if we reject the conclusions of the
\textit{best} approximate theory in some area, then the burden is on us to
explain why.

To illustrate, suppose you believe that quantum computers will never give a
speedup over classical computers for any practical problem. \ Then as an
explanation for your stance, you might assert any of the following:

\begin{enumerate}
\item[(a)] Quantum mechanics is false or incomplete, and an attempt to build a
scalable quantum computer would instead lead to falsifying or extending
quantum mechanics itself.

\item[(b)] There exist polynomial-time \textit{classical} algorithms for
factoring integers, and for all the other problems that admit polynomial-time
quantum algorithms. \ (In complexity terms, the classes $\mathsf{BPP}$ and
$\mathsf{BQP}$ are equal.)

\item[(c)] The \textquotedblleft constant-factor overheads\textquotedblright%
\ involved in building a quantum computer are so large as to negate their
asymptotic advantages, for any problem of conceivable human interest.

\item[(d)] While we don't yet know which of (a)-(c) holds, we can know on some
\textit{a priori} ground that at least one of them has to hold.
\end{enumerate}

The point is that, even if we can't answer every possible shortcoming of a
complexity-theoretic analysis, we can still use it to \textit{clarify the
choices}: to force people to lay some cards on the table, committing
themselves either to a prediction that might be falsified or to a mathematical
conjecture that might be disproved. \ Of course, this is a common feature of
\textit{all} scientific theories, not something specific to complexity theory.
\ If complexity theory is unusual here, it is only in the number of
\textquotedblleft predictions\textquotedblright\ it juggles that could be
confirmed or refuted by mathematical proof (and indeed, \textit{only} by
mathematical proof).\footnote{One other example that springs to mind, of a
scientific theory many of whose \textquotedblleft
predictions\textquotedblright\ take the form of mathematical conjectures, is
string theory.}

\subsection{Future Directions\label{FUTURE}}

Even if the various criticisms of complexity theory don't negate its
relevance, it would be great to address those criticisms head-on---and more
generally, to get a clearer understanding of the relationship between
complexity theory and the real-world phenomena that it tries to explain.
\ Toward that end, I think the following questions would all benefit from
careful philosophical analysis:

\begin{itemize}
\item What is the empirical status of asymptotic claims? \ What sense can we
give to an asymptotic statement \textquotedblleft making
predictions,\textquotedblright\ or being supported or ruled out by a finite
number of observations?

\item How can we explain the empirical facts on which complexity theory
relies: for example, that we rarely see $n^{10000}$\ or $1.0000001^{n}$
algorithms, or that the computational problems humans care about tend to
organize themselves into a relatively-small number of equivalence classes?

\item Short of proof, how do people form intuitions about the truth or
falsehood of mathematical conjectures? \ What \textit{are} those intuitions,
in cases such as $\mathsf{P}\neq\mathsf{NP}$?

\item Do the conceptual conclusions that people sometimes want to draw from
conjectures such as $\mathsf{P}\neq\mathsf{NP}$\ or $\mathsf{BPP}%
\neq\mathsf{BQP}$---for example, about the nature of mathematical creativity
or the interpretation of quantum mechanics---actually depend on those
conjectures being true? \ Are there easier-to-prove statements that would
arguably support the same conclusions?

\item If $\mathsf{P}\neq\mathsf{NP}$, then how have humans managed to make
such enormous mathematical progress, even in the face of the general
intractability of theorem-proving? \ Is there a \textquotedblleft selection
effect,\textquotedblright\ by which mathematicians favor problems with special
structure that makes them easier to solve than arbitrary problems? \ If so,
then what does this structure consist of?
\end{itemize}

In short, I see plenty of scope for the converse essay to this one:
\textquotedblleft Why Computational Complexity Theorists Should Care About
Philosophy.\textquotedblright

\section{Acknowledgments}

I am grateful to Oron Shagrir for pushing me to finish this essay, for helpful
comments, and for suggesting Section \ref{GRUE}; to Alex Byrne for suggesting
Section \ref{WATERFALLS}; to Agust\'{\i}n Rayo for suggesting Section
\ref{OMNI}; and to David Aaronson, Seamus Bradley, Terrence Cole, Michael
Collins, Andy Drucker, Michael Forbes, Oded Goldreich, Bob Harper, Gil Kalai,
Dana Moshkovitz, Jan Arne Telle, Dylan Thurston, Ronald de Wolf, Avi
Wigderson, and Joshua Zelinsky for their feedback.

\bibliographystyle{plain}
\bibliography{thesis}

\end{document}